\documentclass[11pt]{extarticle}
\usepackage[english]{babel}
\usepackage{authblk}
\usepackage{graphicx}
\usepackage{amssymb} 
\usepackage{framed}
\usepackage{mathrsfs}
\usepackage{amsmath}
\usepackage{booktabs}
\usepackage[normalem]{ulem}
\usepackage{indentfirst}
\usepackage{amsmath}
\usepackage{amsthm}
\usepackage{amssymb}
\usepackage{amsfonts}
\usepackage{subfigure}
\usepackage{float}
\usepackage{enumerate}
\usepackage{cite}
\usepackage[utf8]{inputenc}
\usepackage[top=1 in,bottom=1in, left=1 in, right=1 in]{geometry}
\usepackage{algorithm}
\usepackage{algorithmicx}
\usepackage{algpseudocode}
\usepackage{setspace}

\title{Health Assessment and Prognostics Based on Higher Order Hidden Semi-Markov Models}
\author[1]{Ying Liao}
\author[1]{Yisha Xiang}
\author[2]{Min Wang}
\affil[1]{Department of Industrial, Manufacturing \& Systems Engineering, Texas Tech University}
\affil[2]{Department of Mathematics and Statistics, Texas Tech University}
\date{}

\providecommand{\keywords}[1]
{
	\small	
	\textbf{\textit{Keywords:}} #1
}

\begin{document}

\maketitle
\begin{abstract}
This paper presents a new and flexible prognostics framework based on a higher order hidden semi-Markov model (HOHSMM) for systems or components with unobservable health states and complex transition dynamics. The HOHSMM extends the basic hidden Markov model (HMM) by allowing the hidden state to depend on its more distant history and assuming generally distributed state duration. An effective Gibbs sampling algorithm is designed for statistical inference of an HOHSMM. The performance of the proposed HOHSMM sampler is evaluated by conducting a simulation experiment. We further design a decoding algorithm to estimate the hidden health states using the learned model. Remaining useful life (RUL) is predicted using a simulation approach given the decoded hidden states. The practical utility of the proposed prognostics framework is demonstrated by a  case study on NASA turbofan engines. The results show that the HOHSMM-based prognostics framework provides good hidden health state assessment and RUL estimation for complex systems.
\end{abstract}
\newpage
\section{Introduction}
In the past decade, prognostics has emerged as one of the key enablers for industrial systems to become more reliable, operationally available, and economically maintained \cite{sun2012benefits}. Prognostics technologies aim to monitor the performance of a system (or a component), assess the health status, and predict the remaining useful life (RUL). Based on the predicted future performance, informed asset management strategies can be better planned to reduce operational risks and costs. Prognostics has been used for various engineering systems, such as engines \cite{wang2005model,peel2008data,zaidan2016gas}, batteries \cite{saha2008prognostics,zhang2011review}, electronics \cite{rouet2007concept,pecht2009prognostics}, and bearings \cite{li2000stochastic,huang2007residual,qiu2002damage,li1999adaptive}. In this paper, we propose a new and flexible prognostics framework based on a higher order hidden semi-Markov model (HOHSMM) to assess the health state and estimate the RUL of a system using condition monitoring data. We are particularly motivated by applications where the health state cannot be directly observable and the transition dynamics of the hidden health state are complex (e.g., depending on distant past, non-geometric sojourn time in each state). For example, turbofan engines typically have complex failure mechanisms and unobservable health conditions, which can only be inferred from sensor measurements, and health state transitions are often history-dependent, violating the first order Markovian assumption. Novel techniques that can model and predict such complex transition behaviors are needed. 

Prognostics approaches can generally be classified into two categories: model-based approach \cite{li1999adaptive,myotyri2006application,daigle2012model,luo2008model,qiu2002damage,kacprzynski2004predicting,cadini2009model,patil2009precursor}, and data-driven approach \cite{huang2007residual,gebraeel2008neural,wang2004prognosis,wang2007adaptive,yan2004prognostic,swanson2000prognostic,zio2011particle,bunks2000condition,tobon2012data,moghaddass2014integrated}. There are also hybrid models \cite{goebel2006fusing,goebel2007prognostic,saha2008prognostics,kumar2008hybrid} that attempt to combine the strengths of model-based and data-driven approaches by fusing the results from both approaches. The model-based approaches require a good understanding of system physics-of-failure mechanisms. Most of model-based approaches deal with crack, wearing, and corrosion phenomena. For example, Li et al. \cite{li1999adaptive} propose to use the Paris–Erdogan model to predict a bearing's crack propagation and estimate the crack size. Similarly, Paris–Erdogan equation is used to model fatigue crack growth and a stochastic filtering technique is applied for real-time RUL estimation \cite{myotyri2006application}. Daigle and Goebel \cite{daigle2012model} develop a model-based prognostics framework for a centrifugal pump that includes models of the most significant damage progression processes such as impeller wear and bearing wear. Damage processes are characterized by a set of parameterized functions (e.g., erosive wear equation, friction coefficient equation), which describe how damage variables evolve in time. More model-based prognostics methods can be found in \cite{qiu2002damage,kacprzynski2004predicting,luo2008model,cadini2009model,patil2009precursor}. Model-based approaches are built on the knowledge of the processes and failure mechanisms occurring in the system of concern, and therefore the approaches allow for identification of the nature and extent of the fault. However, the limitations of model-based approaches are (1) specific domain knowledge for developing physical models is required but may not always be available, and (2) it is usually a challenging task to create dynamic models representing multiple physical processes occurring in complex systems. 

With the rapid development of sensor technologies, it has become much easier and less costly to obtain condition monitoring data, including operational and environmental loads as well as performance conditions of the monitored system (e.g., temperature, vibration, pressure, voltage, current) \cite{cheng2010sensor}. Advancements in modern sensor instruments have greatly facilitated data-driven prognostics. The monitoring data provide useful information for building a behavior model to characterize the evolution of system performance. Data-driven approaches use several tools, most of which originate from machine learning and statistical domains \cite{peng2010current}. Among different machine learning techniques, neural networks and neuro-fuzzy networks are the most commonly used ones. Huang et al. \cite{huang2007residual} propose a prognostics framework for ball bearings based on self-organizing map (SOM) and neural network. The degradation indicator is extracted using SOM and the residual life is predicted based on back propagation neural networks. Gebraeel and Lawley \cite{gebraeel2008neural} develop a degradation model based on dynamic wavelet neural network to compute and continuously update residual life distributions of partially degraded component using condition-based sensory signals. A neuro-fuzzy network is used to predict the future health state of a gear in \cite{wang2004prognosis}. Furthermore, Wang \cite{wang2007adaptive} develops an adaptive predictor based on neuro-fuzzy approach to forecast the behavior of dynamic systems, where the forecasting is performed by fuzzy logic and the fuzzy system parameters are trained by using neural networks. Various statistical tools have been used for prognostics, including time series analysis models, Kalman and particle filters, and Markov chains. 
Yan et al. \cite{yan2004prognostic} employ a logistic regression model to compute the failure probability given condition variables and then use an autoregressive moving average model to predict future performance and estimate the RUL. Swanson et al. \cite{swanson2000prognostic} use Kalman filter to track the time evolution of a crack in a tensioned steel band. More general than Kalman filter, particle filter can be used to perform nonlinear projection of features, which is exploited for RUL estimation of a mechanical component subject to fatigue crack growth \cite{zio2011particle}. 

Most of the existing strategies estimate RUL by predicting the degradation level (e.g., crack size), which is either directly observable or can be quantified based on sensor signals. In practice, the damage level of many engineering systems (e.g., turbofan engine) cannot be easily quantified due to complex failure mechanisms. Hidden Markov models (HMMs) are commonly used to infer the hidden health state directly from the observed data (e.g., sensor measurements) and predict the RUL. An HMM is defined as a statistical model that is used to represent stochastic processes, where the states are not directly observed but can emit corresponding observations \cite{rabiner1989tutorial}. Bunks et al. \cite{bunks2000condition} illustrate the applications of HMMs by using the Westland helicopter gearbox data set and show that HMMs can provide a natural framework for both health diagnostics and prognostics. Tobon-Mejia et al. \cite{tobon2012data} develop a mixture of Gaussians hidden Markov model (MoG-HMM) to predict RUL of bearings. They use wavelet packet decomposition technique to extract continuous features from the monitoring signals and then use the features as observations to train MoG-HMMs. The learned MoG-HMMs are then exploited to assess the current condition of a bearing and estimate its RUL. However, standard HMMs have two inherent limitations. One is the assumption of first order Markovian dynamics of the hidden state process. The other is that the state duration (i.e., sojourn time) implicitly follows a geometric distribution. The first order assumption can be restrictive as the health state of complex systems usually evolves depending on its more distant history, not just the current state. Moreover, the duration time in one state does not always follow a geometric distribution. To provide a more adequate representation of temporal structure, hidden semi-Markov model (HSMM) extends HMM by assuming that the state duration is generally distributed. Moghaddass and Zuo \cite{moghaddass2014integrated} propose an HSMM-based prognostic framework to assess the health conditions and estimate the RULs for gradually degraded systems. They demonstrate the proposed model by a case study on NASA turbofan engines, where principle component analysis (PCA) is used to extract features from multiple sensor measurements. However, the HSMM in \cite{moghaddass2014integrated} still makes the first order Markovian dynamics assumption. 

In this paper, we propose a new prognostics framework based on HOHSMMs for systems with unobservable health state and complex transition dynamics. In the HOHSMM-based framework, the important features extracted from the monitoring data are used as observations and the underlying health status of the concerned system is represented in the form of hidden states, which evolve depending not only on the current state but also on its more distant history. The sojourn time in each state is generally distributed and is assumed to follow an explicit distribution. We design an effective Gibbs sampling algorithm for model inference and conduct a simulation experiment to evaluate the performance of the proposed HOHSMM sampler. The learned HOHSMM is then exploited to assess the current health state of a functioning system in operation and predict its RUL. Decoding algorithm is developed for health state assessment using the learned model. The RUL is estimated using a simulation approach by generating paths from the current health state to the failure state. Furthermore, we demonstrate the practical utility of the proposed prognostics framework by conducting a case study on NASA turbofan engines. The main contributions of this paper are two-fold. 
\begin{enumerate}[(1)]
	\item Develop a new and advanced HOHSMM-based prognostics framework to assess hidden health state and predict the RUL for complex systems. The proposed HOHSMM includes the HMM and HSMM as two special cases.
	\item Design efficient algorithms for HOHSMM inference, hidden state decoding, and RUL prediction. A Gibbs sampling algorithm is developed for HOHSMM inference and the simulation experiment shows that the designed HOHSMM sampler is effective for learning model parameters from observations. Based on the learned model, a decoding algorithm is developed for hidden health state assessment and an RUL estimation algorithm is developed for prognostics. The case study on NASA turbofan engines shows that the HOHSMM-based prognostics framework provides satisfactory hidden health state assessment and RUL estimation for complex systems.
\end{enumerate} 


The remainder of this paper is organized as follows. Section 2 provides preliminaries on higher order hidden Markov model (HOHMM). In Section 3, we develop an HOHSMM and design an effective sampling algorithm for statistical inference. Section 4 presents the hidden state decoding procedure using the learned model. The RUL is predicted using a simulation approach in Section 5. We conduct a simulation experiment to evaluate the performance of the proposed HOHSMM sampler in Section 6. A case study on NASA turbofan engines is demonstrated in Section 7. Section 8 discusses the concluding remarks and future work.


\section{Preliminaries on Higher Order Hidden Markov Model}

This section provides a brief overview of the Higher Order Hidden Markov Model (HOHMM) by summarizing the main results of \cite{sarkar2018bayesian} and \cite{yang2016bayesian}. Based on the HOHMM in \cite{sarkar2018bayesian}, we develop the HOHSMM. 

An HOHMM consists of two processes: a hidden process \(\{c_t\}\), which evolves according to a higher order Markov chain with discrete state space, and a potentially multivariate observed process \(\{y_t\}\) observed sequentially over a set of discrete time points \(t=1,2,\ldots,T\). HOHMMs extend the idea of basic HMMs by allowing the hidden state sequence \(\{c_t\}\) to depend on its more distant past. An HOHMM of maximal order \(q\) makes the following set of conditional independence assumptions:
\begin{equation}
 p(c_t|c_1,\ldots,c_{t-1})=p\left(c_t|c_{(t-q):(t-1)}\right),
\end{equation}
\begin{equation}
 p(y_t|c_1,\ldots,c_t,y_1,\ldots,y_{t-1})=p(y_t|c_t).
\end{equation}
Note that an HOHMM is said to be of maximal order $q$ if the distribution of $c_t$ only depends on a subset of $\{c_{t-1},\ldots,c_{t-q}\}$. If the distribution of $c_t$ actually varies with the values at all the previous $q$ time points, the HOHMM is considered to be of full order $q$.

\begin{figure}
	\centering
	\subfigure[without allocation variables]{
		\begin{minipage}[t]{1\textwidth}
			\includegraphics[width=4.8in]{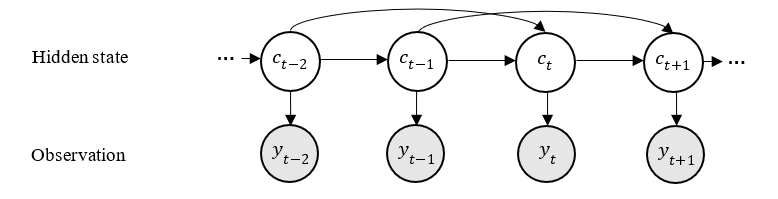}
			\centering
		\end{minipage}
	}
   \subfigure[with allocation variables]{
	\begin{minipage}[t]{1\textwidth}
		\includegraphics[width=4.8in]{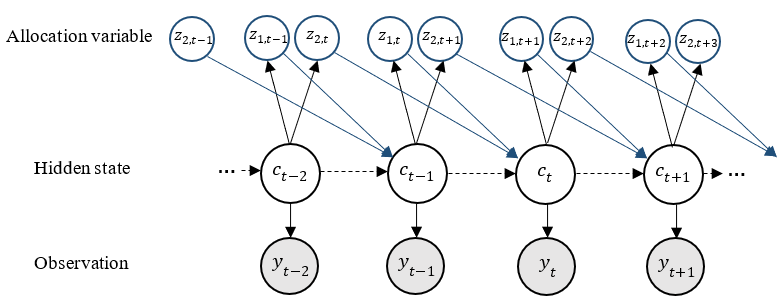}
		\centering
	\end{minipage}
    } 
   \caption{Dependence structure of a second order hidden Markov model} 
   \label{allocation}
\end{figure}

While the HOHMM relaxes the restrictive first order assumption of the basic HMM, it also brings significant dimensionality challenge. For known state space \(\mathscr{C}=\{1,\ldots,C\}\), the transition probabilities are now indexed by $C^q$ different possible values of the lags $c_{(t-q):(t-1)}$ and involve a total number of $(C-1)C^q$ parameters, which increases exponentially with the order $q$. To address this issue, latent allocation variables \(z_{j,t}\) for \(j=1,\ldots,q\) and \(t=q+1,\ldots,T\) are introduced to shrink the total number of parameters. The allocation variable \(z_{j,t}\), taking values from $\{1,\ldots,k_j\}$, is the respective latent class that a particular state of \(c_{t-j}\) is allocated into. The total number of the latent classes \(k_j\) ($1\leq k_j\leq C$) then determines the inclusion of the \(j^{th}\) lag \(c_{t-j}\). If $k_j=1$, it means that \(c_{t-j}\) is not an important lag for \(c_t\). If $k_j>1$ for all $j=1,\ldots,q$, the HOHMM is of full order $q$. Based on the allocation variable \(z_{j,t}\), the hidden states \(\{c_t\}\) are conditionally independent as shown in Figure \ref{allocation}. 

We denote the probability that the \(j^{th}\) lag \(c_{t-j}\) is allocated into latent class $h_j$ by $\pi_{h_j}^{(j)}(c_{t-j})$, i.e., $\pi_{h_j}^{(j)}(c_{t-j})=p(z_{j,t}=h_j|c_{t-j})$. Given the combination of $q$ allocated latent classes $(h_1,\ldots,h_q)$, the state transition probability is denoted by $\lambda_{h_1,\ldots,h_q}(c_t)$ for $c_t=1,\ldots,C$, 
\begin{equation}
\lambda_{h_1,\ldots,h_q}(c_t)=p(c_t|z_{1,t}=h_1,\ldots,z_{q,t}=h_q).
\end{equation}
Then the transition probability can be structured through the following hierarchical formulation
\begin{equation}
 (c_t|z_{j,t}=h_j,j=1,\ldots,q)\sim \text{Mult}(\{1,\ldots,C\},\lambda_{h_1,\ldots,h_q}(1),\ldots,\lambda_{h_1,\ldots,h_q}(C)),
\end{equation}
\begin{equation}
 (z_{j,t}|c_{t-j})\sim \text{Mult}(\{1,\ldots,k_j\},\pi_{1}^{(j)}(c_{t-j}),\ldots,\pi_{k_j}^{(j)}(c_{t-j})).
\end{equation}
The parameters \(\lambda_{h_1,\ldots,h_q}(c_t)\) and \(\pi_{h_j}^{(j)}(c_{t-j})\) are all non-negative and satisfy the constraints:
\begin{enumerate}[(1)]
	\item \(\sum_{c_t=1}^{C}\lambda_{h_1,\ldots,h_q}(c_t)=1\), for each combination \((h_1,\ldots,h_q)\);
	\item \(\sum_{h_j=1}^{k_j}\pi_{h_j}^{(j)}(c_{t-j})=1\), for each pair \((j,c_{t-j})\).
\end{enumerate}
In such a factorization, the number of parameters is reduced to $(C-1)\prod_{j=1}^{q}k_j+C\sum_{j=1}^{q}(k_j-1)$, which is much smaller than $(C-1)C^q$ if $\prod_{j=1}^{q}k_j\ll C^q$.

Marginalizing out the latent class indicators \(z_{j,t}\), the transition probability \(p(c_t|c_{(t-q):(t-1)})\) has an equivalent form as
\begin{equation}
p(c_t|c_{t-j},j=1,\ldots,q)=\sum_{h_1=1}^{k_1}\cdots\sum_{h_q=1}^{k_q}\lambda_{h_1,\ldots,h_q}(c_t)\prod_{j=1}^{q}\pi_{h_j}^{(j)}(c_{t-j}),
\end{equation}
where \(1\leq k_j\leq C\) for all \(j\). Thus, the $r$-step transition probability can be obtained as
\begin{equation}
p(c_{T+r}|\boldsymbol{c}_{(T+1-q):T})=\sum_{c_{T+r-1}}\cdots\sum_{c_{T+1}}p(c_{T+r}|\boldsymbol{c}_{(T+r-q):(T+r-1)})\times\cdots\times p(c_{T+1}|\boldsymbol{c}_{(T+1-q):T}).
\end{equation}

Efficient two-stage Gibbs sampling algorithms have been designed in \cite{sarkar2018bayesian} for HOHMM inference. First, a hierarchical Dirichlet prior is assigned on \(\boldsymbol{\lambda}_{h_1,\ldots,h_q}=\{\lambda_{h_1,\ldots,h_q}(1),\ldots,\lambda_{h_1,\ldots,h_q}(C)\}\),
\begin{equation}
 \boldsymbol{\lambda}_{h_1,\ldots,h_q}\sim \text{Dir}\{\alpha\lambda_0(1),\ldots,\alpha\lambda_0(C)\},\quad\forall(h_1,\ldots,h_q),
 \label{hier1}
\end{equation}
\begin{equation}
 \boldsymbol{\lambda}_0=\{\lambda_0(1),\ldots,\lambda_0(C)\}\sim \text{Dir}(\alpha_0/C,\ldots,\alpha_0/C).
 \label{hier2}
\end{equation}
The dimension of \(\boldsymbol{\pi}_{k_j}^{(j)}(c_{t-j})=\{\pi_{1}^{(j)}(c_{t-j}),\ldots,\pi_{k_j}^{(j)}(c_{t-j})\}\) varies with \(k_j\). Independent priors on the \(\boldsymbol{\pi}_{k_j}^{(j)}(c_{t-j})\)'s are assigned as
\begin{equation}
 \boldsymbol{\pi}_{k_j}^{(j)}(c_{t-j})=\{\pi_{1}^{(j)}(c_{t-j}),\ldots,\pi_{k_j}^{(j)}(c_{t-j})\}\sim \text{Dir}(\gamma_j,\ldots,\gamma_j), \quad\forall (j,c_{t-j}).
\end{equation}
Finally, the following independent priors are assigned on \(k_j\)'s
\begin{equation}
 p_{0,j}(k)\propto \text{exp}(-\varphi jk), \quad j=1,\cdots,q,\quad k=1,\cdots,C,
\end{equation}
where \(\varphi>0\). The prior $p_{0,j}$ assigns increasing probabilities to smaller values of $k_j$ as the lag $j$ becomes more distant, reflecting the natural belief that increasing lags have diminishing influence on the distribution of $c_t$. The generic form of the emission distribution is expressed as $p(y_t|c_t,\boldsymbol{\theta})=f(y_t|\boldsymbol{\theta}_{c_t})$, where \(\boldsymbol{\theta}=\{\boldsymbol{\theta}_c: c=1,\ldots,C\}\) represents parameters indexed by the hidden states.

The joint distribution of \(\boldsymbol{y}=\{y_t: t=1,\ldots,T\}\), \(\boldsymbol{c}=\{c_t: t=q+1,\ldots,T\}\) and \(\boldsymbol{z}=\{z_{j,t}: t=q+1,\ldots,T,j=1,\ldots,q\}\) admits the following factorization
\begin{equation}
\begin{aligned}
 p(\boldsymbol{y},\boldsymbol{c},\boldsymbol{z}|\boldsymbol{\lambda_k},\boldsymbol{\pi_k},\boldsymbol{k},\boldsymbol{\theta})&=\prod_{t=q+1}^{T}\left\{p(c_t|\boldsymbol{\lambda}_{\boldsymbol{z}_t})\prod_{j=1}^{q}p(z_{j,t}|w_{j,t},\boldsymbol{\pi}^{(j)}_{k_j},k_j)\right\}\prod_{t=1}^{T}f(y_t|\boldsymbol{\theta}_{c_t})
 \\&=p(\boldsymbol{y}|\boldsymbol{c},\boldsymbol{\theta})p(\boldsymbol{c}|\boldsymbol{z},\boldsymbol{\lambda_k},\boldsymbol{k})p(\boldsymbol{z}|\boldsymbol{w},\boldsymbol{\pi_k},\boldsymbol{k}),
\end{aligned}
\end{equation}
where \(w_{j,t}=c_{t-j}\), representing the history state of $c_t$. The conditional independence relationships encoded in the factorization are used in deriving MCMC algorithms to draw samples from the posteriors. Detailed sampling algorithms are referred to \cite{sarkar2018bayesian}.

\section{Higher Order Hidden Semi-Markov Model}
In this paper, we extend an HOHMM to a higher order hidden semi-Markov model (HOHSMM), where the hidden state sequence is governed by a semi-Markov chain. The HOHSMM is more flexible since it incorporates additional temporal structure by allowing the state duration to be generally distributed, rather than implicitly following a geometric distribution as in an HOHMM.

\subsection{Model development}
We first give a brief description of the HSMM and then develop the HOHSMM. There exist several specific models of HSMM which make different assumptions regarding the dependence between state transition and duration, for example, residential time HMMs and explicit duration HMMs \cite{yu2010hidden}. A residential time HMM assumes that the current state and its duration time are determined by the previous state, and independent to the duration of the previous state. An explicit duration HMM assumes that a transition to the current state is independent to the duration of the previous state and the duration is only conditional on the current state. We consider the explicit duration setting in our HOHSMM. Both HSMMs and HOHSMMs with explicit duration exclude state self-transitions because the duration distribution can not fully capture a state's possible duration time if self-transitions are allowed.

\begin{figure}[h]
	\centering
	\includegraphics[width=6in]{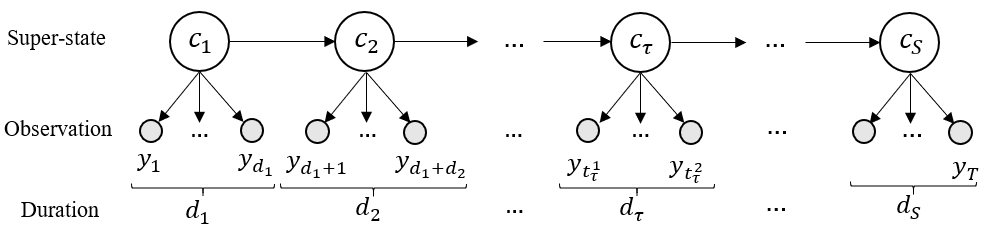}
	\caption{An explicit duration HMM}
	\label{hsmm}
\end{figure}

An explicit duration HMM assumes that the underlying stochastic process is governed by a semi-Markov chain \cite{yu2010hidden}. Each state has a variable duration that follows an explicit state-specific distribution and a number of corresponding observations are produced while in the state (illustrated in Figure \ref{hsmm}). 
The observation sequence $\{y_t: t=1,\ldots,T\}$ is produced segmentally from the emission distribution $f(y|\theta_{c_{\tau}})$ indexed by the hidden super-state sequence $\{c_{\tau}:\tau=1,\ldots,S\}$, where $S$ is the number of segments. Observations are assumed to be collected discretely by a unit time, and therefore the number of observations produced in each super-state represents the state duration. For the $\tau^{th}$ segment, the state duration is denoted by $d_{\tau}$ and $\boldsymbol{y}_{t_{\tau}^1:t_{\tau}^2}$ denotes the produced observations, where $t_{\tau}^1=\sum_{\psi<\tau}d_{\psi}+1$, $t_{\tau}^2=\sum_{\psi\leq\tau}d_{\psi}$. In the last segment, the observations may be truncated, and we have $t_S^2=\min\left\{\sum_{\psi\leq S}d_{\psi},T\right\}$.

\begin{figure}[h]
	\centering
	\includegraphics[width=5.7in]{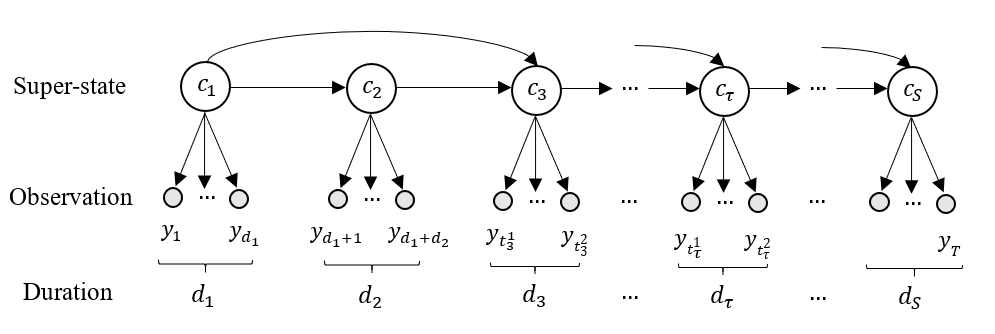}
	\caption{A second order HSMM}
	\label{hohsmm_2order}
\end{figure}

In the proposed HOHSMM with the explicit duration setting, the hidden super-state sequence is assumed to be governed by a higher order Markov chain and the state duration follows an explicit distribution (e.g., Poisson distribution), denoted by  $g(d|\xi_{c_{\tau}})$ with the parameters indexed by the specific hidden super-state $c_{\tau}$. An explicit-duration HOHSMM of maximal order $q$ is constructed as follows
\begin{equation*}
p(c_{\tau}|c_1,\ldots,c_{\tau-1})=p(c_{\tau}|c_{\tau-q},\ldots,c_{\tau-1}),
\quad\tau=q+1,\ldots,S,\\
\end{equation*}
\begin{equation*}
d_{\tau}\sim g(d|\xi_{c_{\tau}}),\quad\tau=1,\ldots,S,
\end{equation*}
\begin{equation*}
\boldsymbol{y}_{t_{\tau}^1:t_{\tau}^2}\stackrel{\text{iid}}{\sim} f(y|\theta_{c_{\tau}}),\quad t_{\tau}^1=\sum_{\psi<\tau}d_{\psi}+1,\quad t_{\tau}^2=\sum_{\psi\leq\tau}d_{\psi}.
\end{equation*}
Figure \ref{hohsmm_2order} illustrates a second order HSMM. In this example, the distribution of the hidden super-state $c_{\tau}$ depends on its previous two states $c_{\tau-1}$ and $c_{\tau-2}$, and the duration time in each super-state is generally distributed, following an explicit state-specific distribution.

The hierarchical Dirichlet prior assigned for transition distribution parameter in Equations (\ref{hier1}) and (\ref{hier2}) does not exclude self-transitions. In order to exclude self-transitions in the super-state sequence for an HOHSMM, a modified hierarchical Dirichlet prior is assigned as  \cite{johnson2013bayesian}
\begin{equation}
\boldsymbol{\lambda}_{i,h_2,\ldots,h_q}=\{\lambda_{i,h_2,\ldots,h_q}(1),\ldots,\lambda_{i,h_2,\ldots,h_q}(C)\}\sim \text{Dir}\{\alpha\lambda_0(1),\ldots,\alpha\lambda_0(C)\},\quad\forall(i,h_2,\ldots,h_q),
\end{equation}
\begin{equation}
\boldsymbol{\lambda}_0=\{\lambda_0(1),\cdots,\lambda_0(C)\}\sim \text{Dir}(\alpha_0/C,\cdots,\alpha_0/C),
\end{equation}
\begin{equation}
\bar\lambda_{i,h_2,\ldots,h_q}(i^{\prime}):=\frac{\lambda_{i,h_2,\ldots,h_q}(i^{\prime})}{1-\lambda_{i,h_2,\ldots,h_q}(i)}(1-\delta_{ii^{\prime}}),\quad \delta_{ii^{\prime}}=\begin{cases}
1 &\text{if $i=i^{\prime}$,}\\
0 &\text{otherwise.}
\end{cases}
\label{self}
\end{equation}
Equation (\ref{self}) ensures that the self-transition probabilities are zeros. Note that $i$ is the latent class the hidden super-state $c_{\tau-1}$ (the immediate precedent super-state of $c_{\tau}$) is allocated into and $i^\prime$ is the state of $c_{\tau}$. Therefore, to have a valid comparison between $i$ and $i^\prime$ and exclude self-transitions, each possible state of $c_{\tau-1}$ must be allocated to a distinct latent class. In other words, each state of $c_{\tau-1}$ has its own latent class. 
To do so, we let $k_1=C$ and $\boldsymbol{\pi}_C^{(1)}(c_{\tau-1})=\{\pi_{1}^{(1)}(c_{\tau-1}),\ldots,\pi_{C}^{(1)}(c_{\tau-1})\}$, where $\pi_i^{(1)}(c_{\tau-1})=\delta_{i,c_{\tau-1}}$, $\forall i=1,\ldots,C$ and $\tau=q+1,\ldots,S$. For the remaining lags, the independent priors on the allocation distribution $\boldsymbol{\pi_k}$ are assigned as 
\begin{equation}
\boldsymbol{\pi}_{k_j}^{(j)}(c_{\tau-j})=\{\pi_{1}^{(j)}(c_{\tau-j}),\ldots,\pi_{k_j}^{(j)}(c_{\tau-j})\}\sim \text{Dir}(\gamma_j,\cdots,\gamma_j), \quad\forall (j,c_{\tau-j}),\quad j=2,\ldots,q.
\end{equation}
The transition probability is modeled as
\begin{equation}
p(c_{\tau}|c_{\tau-j},j=1,\ldots,q)=\sum_{i=1}^{C}\sum_{h_2=1}^{k_2}\cdots\sum_{h_q=1}^{k_q}\bar\lambda_{i,h_2,\ldots ,h_q}(c_{\tau})\;\pi_{i}^{(1)}(c_{\tau-1})\prod_{j=2}^{q}\pi_{h_j}^{(j)}(c_{\tau-j}).
\label{transitions}
\end{equation}
Similarly, by introducing latent allocation variables \(z_{j,\tau}\) for \(j=1,\ldots,q\) and \(\tau=q+1,\ldots,S\) with $z_{1,\tau}=c_{\tau-1}$,  the hidden super-states $\{c_{\tau}\}$ are conditionally independent and the model can be represented through the following hierarchical formulation
\begin{equation}
(c_{\tau}|z_{1,\tau}=i,z_{j,\tau}=h_j,j=2,\ldots,q)\sim \text{Mult}(\{1,\ldots,C\},\bar\lambda_{i,h_2,\ldots,h_q}(1),\ldots,\bar\lambda_{i,h_2,\ldots,h_q}(C)),
\end{equation}
\begin{equation}
(z_{j,\tau}|c_{\tau-j})\sim \text{Mult}(\{1,\ldots,k_j\},\pi_{1}^{(j)}(c_{\tau-j}),\ldots,\pi_{k_j}^{(j)}(c_{\tau-j})), \quad\forall j=1,\ldots,q.
\end{equation}

\subsection{Model inference}
We use the MCMC sampling method for explicit-duration HOHSMM inference.
The sampler is designed based on the two-stage Gibbs sampling algorithms for HOHMM \cite{sarkar2018bayesian}. There are additional challenges due to explicit temporal structure, excluding self-transitions, and multiple observed trajectories in the training data. 

The first challenge is brought by incorporating explicit temporal structure (i.e., duration distribution), which requires additional sampling to determine the number of segments (i.e., the number of hidden super-states) and the duration time in each state. Existing sampling inference methods for HSMMs often use a message-backwards and sample-forwards technique to address this problem \cite{johnson2013bayesian}. However, it is not applicable for an HOHSMM since the backwards messages are extremely difficult to define and compute when higher order transitions present. The reversible jump MCMC provides a statistical inference strategy for Bayesian model determination, where the dimensionality of the parameter vector is typically not fixed (e.g, the multiple change-point problem for Poisson processes) \cite{green1995reversible}. However, it cannot be used to sample change-points of a sequence in an HOHSMM since there is no appropriate mechanism to update the hidden super-states affected by the moves of change-points (e.g., birth of a change-point, death of a change-point). The second challenge is brought by excluding self-transitions. A Dirichlet distribution is assigned as the conjugate prior for transition probability parameters. However, the conjugacy does not exist after setting self-transition probabilities to zeros. A mechanism to recover the conjugacy for updating transition probability parameters is needed. In addition, in many real-world applications, several identical units are typically monitored at the same time to collect sensor data. How to leverage all information provided by multiple observed trajectories (i.e., observation sequences) instead of using just one sequence is the third challenge. We address these difficulties in the following two sections.

\subsubsection{Update segmentation}
We denote $P$ run-to-failure observation sequences by $\boldsymbol{y}^{(1:P)}$ and the $p^{th}$ observation sequence by $\boldsymbol{y}^{(p)}=\{y^{(p)}_t: t=1,\ldots,T_p\}$, where $T_p$ is the observed length and $p=1,\ldots,P$. These sequences are assumed to be independent. To address the first challenge, we introduce a jump size threshold ($\alpha$) to identify change-points. For each observation sequence, if the Euclidean distance between a point and its immediate previous point is greater than $\alpha$, this point is identified as a change-point. The prior of $\alpha$ is assigned to be a uniform distribution with support $(\alpha_{\text{min}}, \alpha_{\text{max}})$, where $\alpha_{\text{min}}$ and $\alpha_{\text{max}}$ are the $5^{th}$ and $95^{th}$ percentile values obtained 
\begin{figure}[H]
	\centering
	\subfigure[Identify change-points given jump size threshold $\alpha$. A red cross indicates a change-point that is detected if the difference (absolute value) between it and its previous observation is larger than $\alpha$. Nine change-points are identified and the observation sequence is segmented accordingly as presented by vertical dashed black lines. Clustering labels are derived by clustering the mean values of observations in these ten segments.]{
		\begin{minipage}{6.1in}
			\includegraphics[width=6.1in]{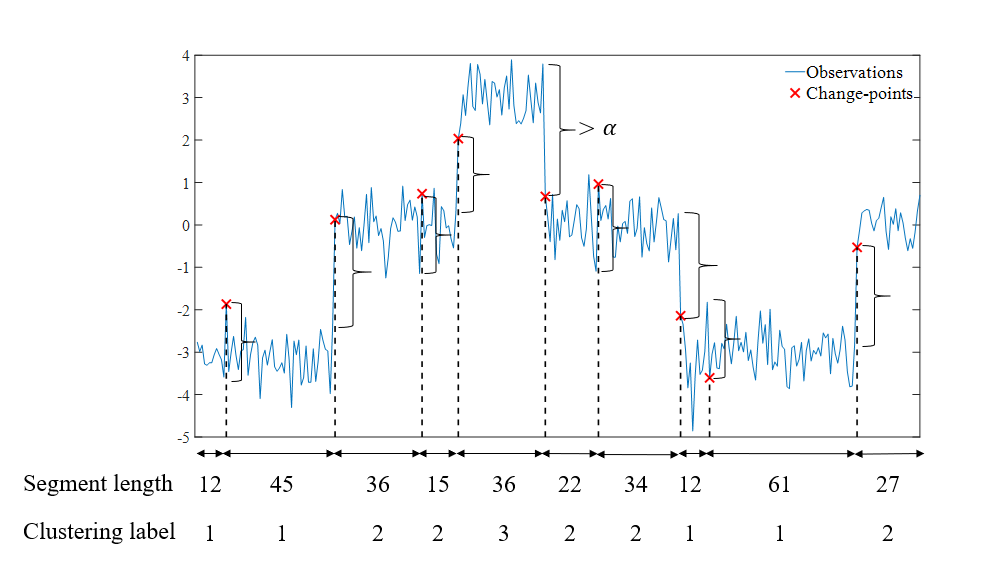}
		\end{minipage}
		\label{jump1}
	}
	
	\subfigure[Segmentation and hidden super-states initialization after clustering and merging processes. If two adjacent segments have the same clustering label, merge these two segments and use the clustering label as the initialized hidden super-state.]{
		\begin{minipage}{6.1in}
			\includegraphics[width=6.1in]{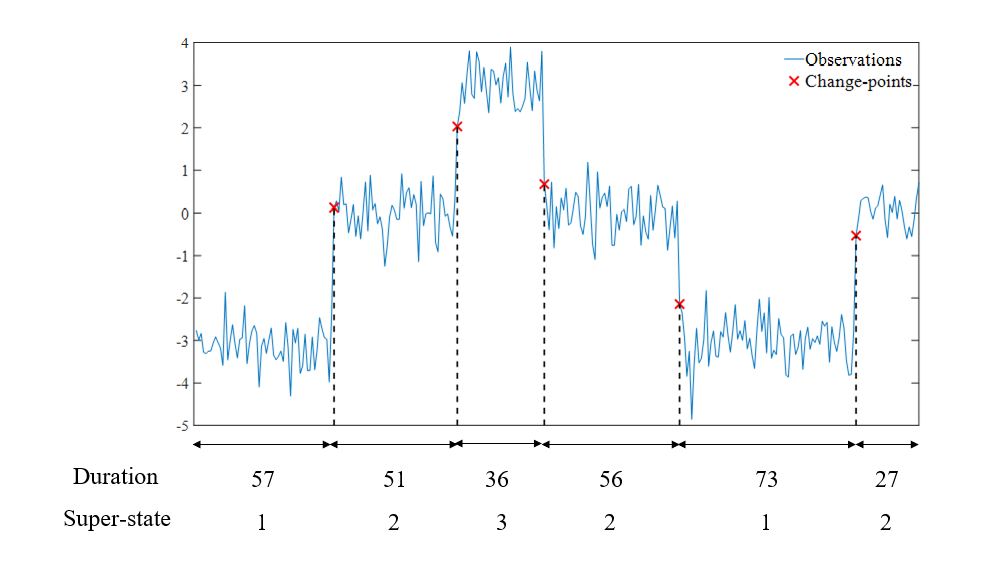}
		\end{minipage}
		\label{jump2}
	}
	\caption{Illustration for updating segmentation and initializing hidden super-states given jump size threshold $\alpha$ (using one-dimensional observation as an example)}
	\label{jump_0}
	
\end{figure} 
\noindent from the distances between two adjacent observed data points in all observation sequences, respectively. We then update the segmentation of the observation sequences and initialize the hidden super-state sequences iteratively by sampling the jump size threshold $\alpha$. 

In each iteration of the HOHSMM sampler, we propose a new threshold $\alpha$ from  $U(\alpha_{\text{min}}, \alpha_{\text{max}})$. For each observation sequence, we mark change-points based on the computed distances (illustrated in Figure \ref{jump1}) and the sequence is segmented accordingly. After the initial segmentation, we compute the center of the observed data points for each segment and label the segments by clustering the centers. The hidden super-states are initialized by using the clustering labels. To exclude self-transitions, if two adjacent segments have the same clustering label, we merge these two segments. For example, the first two segments in Figure \ref{jump1} share the same label 1, and these two segments are merged into one. After the clustering and merging processes, we obtain the final segmentation and the initialized hidden super-states of an observation sequence for a given jump size threshold (illustrated in Figure \ref{jump2}). Based on the segmentation results, we also obtain the number of segments and the state duration for each observation sequence, denoted as $S_p$ and $\boldsymbol{d}^{(p)}$, respectively. 

The hidden super-state sequence $\boldsymbol{c}^{(p)}$, latent allocation variables \(\boldsymbol{z}^{(p)}\), and other parameters $\boldsymbol{k}$, $\boldsymbol{\pi}_{\boldsymbol{k}}$, $\boldsymbol{\lambda_k}$, $\boldsymbol{\bar\lambda}_{\boldsymbol{k}}$, $\boldsymbol{\lambda}_0$, \(\boldsymbol{\theta}\) are updated using the two-stage Gibbs sampling algorithm for the HOHSMM. The first stage is to identify the important lags by sampling $\boldsymbol{k}$ from the posterior. Given the determined $\boldsymbol{k}$, we collect the samples of other parameters in the second stage. The obtained samples will be used to compute the acceptance probability for updating jump size threshold $\alpha$. In general MCMC sampling, the acceptance probability can be computed as
\begin{equation}
\label{rate}
\min\{1,(\text{likelihood ratio})\times(\text{prior ratio})\times(\text{proposal ratio})\}.
\end{equation}
Since the prior of $\alpha$ is a uniform distribution and $\alpha$ is also proposed from the uniform distribution, it is obvious that the prior ratio and the proposal ratio are equal to 1. The posterior mean of the likelihood can be approximated using the obtained samples \cite{ando2010bayesian}, which is provided as 
\begin{equation}
L_{\alpha}=\frac{1}{N}\sum_{j=1}^{N} f\left(\boldsymbol{y}^{(1:P)}|\boldsymbol{c}^{(1:P),j},\boldsymbol{\theta}^{j},\alpha\right)=	\frac{1}{N}\sum_{j=1}^{N}\left[\prod_{p=1}^{P}\prod_{\tau=1}^{S_p} f\left(\boldsymbol{y}^{(p)}_{t_{\tau}^1:t_{\tau}^2}|\boldsymbol{\theta}^{j}_{c_{\tau}^{(p),j}}\right)\right],
\label{log-likeli}
\end{equation}
where $\{\boldsymbol{\theta}^1,\ldots,\boldsymbol{\theta}^N,\boldsymbol{c}^{(1:P),1},\ldots,\boldsymbol{c}^{(1:P),N}\}$ is a set of posterior samples generated from their posterior distributions. 

Given all the collected samples of $\alpha$, the most likely jump size threshold $\alpha^*$ is determined by computing the average value of the samples after burn-in. We then use $\alpha^*$ to update segmentation and repeat the two-stage Gibbs sampling process to obtain the final segmentation and samples for all parameters. Given an explicit distribution $g(d|\xi_{c})$ for each super-state's duration, the MLEs $\{\hat\xi_c\}$ for parameters $\{\xi_{c}: c=1,\ldots,C\}$ can be easily obtained using the final segmentation result.

\subsubsection{The two-stage Gibbs sampling algorithm for HOHSMMs}
Given the segmentation, we modify the two-stage Gibbs sampling algorithm in \cite{sarkar2018bayesian} to determine the values of $\boldsymbol{k}$, $\boldsymbol{\pi}_{\boldsymbol{k}}$, $\boldsymbol{\lambda_k}$, $\boldsymbol{\bar\lambda}_{\boldsymbol{k}}$, $\boldsymbol{\lambda}_0$, \(\boldsymbol{\theta}\), \(\boldsymbol{z}\) and $\boldsymbol{c}$ in the HOHSMM. The first stage is to determine the values of $\boldsymbol{k}=\{k_1,\ldots,k_q\}$, which is the important lag indicator. Given determined values of $\boldsymbol{k}$, we update other model parameters $\boldsymbol{\pi}_{\boldsymbol{k}}$, $\boldsymbol{\lambda_k}$, $\boldsymbol{\bar\lambda}_{\boldsymbol{k}}$, $\boldsymbol{\lambda}_0$, \(\boldsymbol{\theta}\), latent allocation variables \(\boldsymbol{z}\), and hidden super-state sequence \(\boldsymbol{c}\) in the second stage. The joint distribution of \(\boldsymbol{y}^{(p)}=\{y^{(p)}_t: t=1,\ldots,T_p\}\), \(\boldsymbol{c}^{(p)}=\{c_{\tau}^{(p)}: \tau=q+1,\ldots,S_p\}\) and \(\boldsymbol{z}^{(p)}=\{z^{(p)}_{j,\tau}: \tau=q+1,\ldots,S_p,j=1,\cdots,q\}\) for the $p^{th}$ sequence can be presented as
\begin{equation}
p(\boldsymbol{y}^{(p)},\boldsymbol{c}^{(p)},\boldsymbol{z}^{(p)}|\boldsymbol{\bar\lambda_k},\boldsymbol{\pi_k},\boldsymbol{k},\boldsymbol{d}^{(p)},\boldsymbol{\theta})=\prod_{\tau=q+1}^{S_p}\left\{p(c^{(p)}_{\tau}|\boldsymbol{\bar\lambda}_{\boldsymbol{z}^{(p)}_{\tau}})\prod_{j=1}^{q}p(z^{(p)}_{j,\tau}|w^{(p)}_{j,\tau},\boldsymbol{\pi}^{(j)}_{k_j},k_j)\right\}\prod_{\tau=1}^{S_p}f(\boldsymbol{y}^{(p)}_{t_{\tau}^1:t_{\tau}^2}|\boldsymbol{\theta}_{c^{(p)}_{\tau}}),
\label{joint_distribution}
\end{equation}
where \(w^{(p)}_{j,\tau}=c^{(p)}_{\tau-j}\), and $f(\boldsymbol{y}^{(p)}_{t_{\tau}^1:t_{\tau}^2}|\boldsymbol{\theta}_{c_{\tau}})=\prod_{i=t_{\tau}^1}^{t_{\tau}^2}f(y^{(p)}_i|\boldsymbol{\theta}_{c^{(p)}_{\tau}})$, $t_{\tau}^1=\sum_{\psi<\tau}d^{(p)}_{\psi}+1$, $t_{\tau}^2=\sum_{\psi\leq\tau}d^{(p)}_{\psi}$. 
To address the third challenge of multiple trajectories, we use the joint distribution of all observation sequences. Based on the assumption that all sequences are independent, the joint distribution can be obtained as follows  
\begin{equation}
p(\boldsymbol{y}^{(1:P)},\boldsymbol{c}^{(1:P)},\boldsymbol{z}^{(1:P)}|\boldsymbol{\bar\lambda_k},\boldsymbol{\pi_k},\boldsymbol{k},\boldsymbol{d}^{(1:P)},\boldsymbol{\theta})=\prod_{p=1}^P p(\boldsymbol{y}^{(p)},\boldsymbol{c}^{(p)},\boldsymbol{z}^{(p)}|\boldsymbol{\bar\lambda_k},\boldsymbol{\pi_k},\boldsymbol{k},\boldsymbol{d}^{(p)},\boldsymbol{\theta}).
\end{equation} 
The conditional independence relationships encoded in the joint distribution are used in deriving the two-stage Gibbs sampling algorithm for HOHSMMs. 

Specifically, in the first stage, we identify important lags and the corresponding number of latent classes by sampling \(\boldsymbol{k}\). In this stage, we use an approximated model which forces hard allocation of $z_{j,\tau}$'s instead of soft allocation. Hard allocation means that, partition the state space into $k_j$ clusters for the $j^{th}$ lag, then each cluster corresponds to its own latent class. In other words, each state is allocated into one class with probability 1. For example, partition the states \(\{1,2,3,4,5,6\}\) into $k_j=2$ clusters with \(\mathscr{C}_{j,1}=\{1,2,3\}\) and \(\mathscr{C}_{j,2}=\{4,5,6\}\) for the $j^{th}$ lag, hard allocation means that $c_{\tau-j}=1,2$, and 3 will be allocated to the first latent class and $c_{\tau-j}=4,5$, and 6 will be allocated to the second one with probability 1. In soft allocation, one state can be allocated into several possible classes with specific probabilities. The mixture probabilities in the approximated model are denoted by $\tilde{\boldsymbol{\pi}}_{\boldsymbol{k}}$, indicating hard clustering while $\boldsymbol{\pi}_{\boldsymbol{k}}$ indicates soft allocation. 

Based on the approximated model, samples of the parameters are drawn from their respective conditional posteriors following the pre-specified order. We first examine the posteriors of the transition distributions \(\boldsymbol{\lambda_k}\) and \(\boldsymbol{\lambda}_0\). There exist computational machineries of sampling from the posteriors in hierarchical Dirichlet process (HDP) models \cite{Hierarchical2006}. In the HOHMM, the Dirichlet distribution is the conjugate prior of transition probability parameter, so it is straightforward to update the parameters \(\boldsymbol{\lambda_k}\). However, in our HOHSMM, the method used to exclude self-transitions makes the model not fully conjugate. Specifically, let $n_{i,h_2,\ldots,h_q}(c)=\sum_p\sum_{\tau}1\{z^{(p)}_{1,\tau}=i,z^{(p)}_{2,\tau}=h_2,\ldots,z^{(p)}_{q,\tau}=h_q,c^{(p)}_{\tau}=c\}$, which counts the number of transitions from the latent allocation classs $(i,h_2,\ldots,h_q)$ to state $c$ among all observation sequences where $i=1,\ldots,C$ and $h_j=1,\ldots,k_j$ for $j=2,\ldots,q$. Because of no self-transitions, we have $n_{i,h_2,\ldots,h_q}(i)=0$. We consider the posterior distribution of $\boldsymbol{\lambda}_{1,h_2,\ldots,h_q}=\{\lambda_{1,h_2,\ldots,h_q}(1),\lambda_{1,h_2,\ldots,h_q}(2),\ldots,\lambda_{1,h_2,\ldots,h_q}(C)\}$,

\begin{equation}
\begin{aligned}
p(\boldsymbol{\lambda}_{1,h_2,\ldots,h_q}|\boldsymbol{\lambda}_0,\boldsymbol{c},\boldsymbol{z})\propto&\left[\lambda_{1,h_2,\ldots,h_q}(1)\right]^{\alpha\lambda_0(1)-1}\left[\lambda_{1,h_2,\ldots,h_q}(2)\right]^{\alpha\lambda_0(2)-1}\cdots\left[\lambda_{1,h_2,\ldots,h_q}(C)\right]^{\alpha\lambda_0(C)-1}\\
&\times\left(\frac{\lambda_{1,h_2,\ldots,h_q}(2)}{1-\lambda_{1,h_2,\ldots,h_q}(1)}\right)^{n_{1,h_2,\ldots,h_q}(2)}\cdots\left(\frac{\lambda_{1,h_2,\ldots,h_q}(C)}{1-\lambda_{1,h_2,\ldots,h_q}(1)}\right)^{n_{1,h_2,\ldots,h_q}(C)}.
\end{aligned}
\end{equation}
Because of the extra $\frac{1}{1-\lambda_{1,h_2,\ldots,h_q}(1)}$ terms from the likelihood by excluding self-transitions, we cannot reduce this expression to the Dirichlet form over the components of $\boldsymbol{\lambda}_{1,h_2,\ldots,h_q}$. Therefore, the model is not fully conjugate and new posteriors need to be derived. To recover conjugacy, we introduce auxiliary variables $\{\rho_s\}_{s=1}^n$, where $n=\sum_{c=1}^{C}n_{i,h_2,\ldots,h_q}(c)$. Each $\rho_s$ is independently drawn from a geometric distribution with specific success parameter $1-\lambda_{i,h_2,\ldots,h_q}(i)$ \cite{johnson2013bayesian}. We adjust the sampling algorithm by updating transition parameters $\boldsymbol{\lambda}_{i,h_2,\ldots,h_q}=\{\lambda_{i,h_2,\ldots,h_q}(1),\ldots,\lambda_{i,h_2,\ldots,h_q}(i),\ldots,\lambda_{i,h_2,\ldots,h_q}(C)\}$ from the posterior distribution
\begin{equation*}
	\text{Dir}\{\alpha\lambda_0(1)+\\n_{i,h_2,\ldots,h_q}(1),\ldots,\alpha\lambda_0(i)+\sum_{s=1}^{n}\rho_s,\ldots,\alpha\lambda_0(C)+n_{i,h_2,\ldots,h_q}(C)\}.
\end{equation*}
Then we compute $\boldsymbol{\bar\lambda}_{\boldsymbol{k}}$ from Equation (\ref{self}) and update $\boldsymbol{\lambda}_0$.

Since the observation sequences are independent, \(\boldsymbol{c}^{(p)}\) and \(\boldsymbol{z}^{(p)}\) are updated sequence by sequence. For each sequence,  \(\boldsymbol{c}^{(p)}\) and \(\boldsymbol{z}^{(p)}\) are sampled by applying a Metropolis-Hastings step and using simulated annealing to facilitate the convergence. The full conditionals of $\boldsymbol{\theta}$ will depend on the choice of the emission distribution. Finally, a stochastic search variable selection (SSVS) method \cite{george1997approaches} is used to sample $\boldsymbol{k}$ from their posteriors and $\tilde{\boldsymbol{\pi}}_{\boldsymbol{k}}$ are updated by the latent allocation cluster mapping. In the first stage, important lags can be determined and the number of latent classes for each important lag can be derived based the samples of \(\boldsymbol{k}\).

The second stage, given the important lag inclusion result, is to sample parameters $\boldsymbol{\pi}_{\boldsymbol{k}}$, $\boldsymbol{\lambda_k}$, $\boldsymbol{\bar\lambda}_{\boldsymbol{k}}$, $\boldsymbol{\lambda}_0$, \(\boldsymbol{\theta}\), \(\boldsymbol{z}^{(p)}\) and $\boldsymbol{c}^{(p)}$ iteratively. Given the segmentation, the elements of \(\boldsymbol{c}^{(p)}\), \(\boldsymbol{z}^{(p)}\) and \(\boldsymbol{\pi_k}\) have either multinomial or Dirichlet full conditionals and can be straightforwardly updated. Sampling of \(\boldsymbol{\lambda_k}\), $\boldsymbol{\bar\lambda}_{\boldsymbol{k}}$, \(\boldsymbol{\lambda}_0\) and emission parameters \(\boldsymbol{\theta}\) is the same as described in first stage. Details of the HOHSMM inference method are summarized in Algorithm \ref{algorithm1}.

\section{Health States Decoding}
\begin{figure}[h]
	\centering
	\includegraphics[width=5.5in]{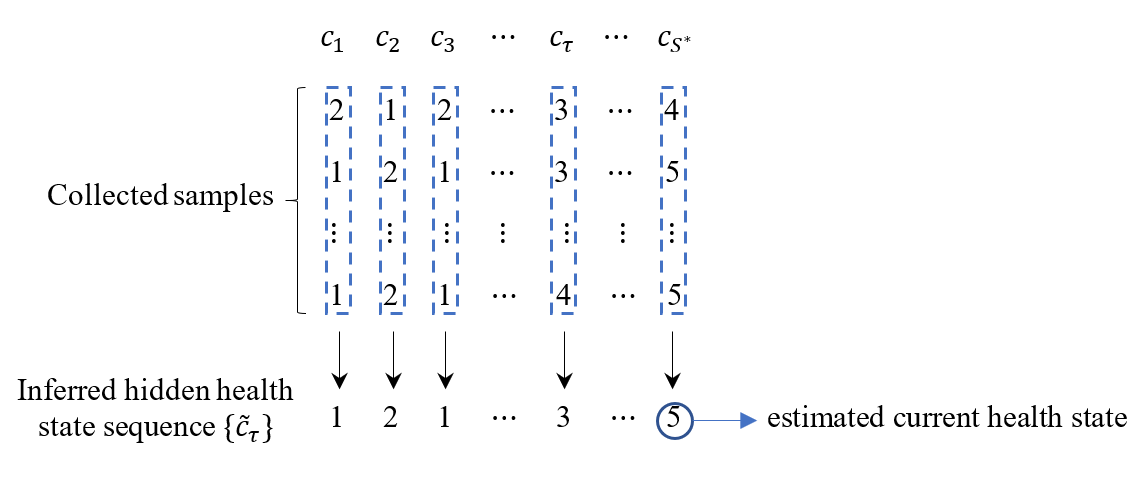}
	\caption{Illustration for determining the hidden health states}
	\label{decoding_state}
\end{figure}

The ultimate purpose of a prognostics framework is to assess the current condition of a system (or component) and to make inferences regarding the remaining useful life (RUL). In this section, we first present how to use the HOHSMM-based prognostics framework to decode the hidden super-states. For an operating system with observation sequence $\boldsymbol{y}$, Equation (\ref{joint_distribution}) provides the joint distribution of $\boldsymbol{y}$, $\boldsymbol{c}$ and $\boldsymbol{z}$ 
\begin{algorithm}[H]
	\caption{Explicit-duration HOHSMM Sampler}
	\label{algorithm1}
	\begin{algorithmic}[1]
		\Require  Observation sequences $\{\boldsymbol{y}^{(p)}: p=1,\ldots,P\}$ and sample size $l$.
		\State \textbf{Initialization}:
		\State \quad Compute distances between two adjacent data points in all sequences $\boldsymbol{y}^{(p)}$ and use the $5^{th}$\par 
		and $95^{th}$ percentile values as the lower bound and upper bound of the support $(\alpha_{\text{min}}, \alpha_{\text{max}})$.		
		\State  \quad Set initial likelihood value: $L_0\leftarrow e^{-10^{10}}$.
		\For{$v=1$ to $l$}
		\State Sample $\alpha_{v}\sim U(\alpha_{\text{min}}, \alpha_{\text{max}})$.
		\State \textbf{Update segmentation}:
		\State \quad For each $\boldsymbol{y}^{(p)}$, identify change-points given $\alpha_{v}$. Compute the center of the observed data\par
		\quad points for each segment and initialize hidden super-state sequence $\{c^{(p)}_{\tau}\}$ by clustering\par
		\quad the centers. Merge adjacent segments that have the same label and derive the number\par 
		\quad of segment $S_p$ and duration times $\boldsymbol{d}^{(p)}$, where $\tau=1,\ldots,S_p$, $p=1,\ldots,P$.
		\State \textbf{Stage 1} (Determine $\boldsymbol{k}$):
		\State \quad Update $\boldsymbol{\lambda_k}$:
		\State \quad\quad Let $n_{i,h_2,\ldots,h_q}(c)=\sum_p\sum_{\tau}1\{z^{(p)}_{1,\tau}=i,z^{(p)}_{2,\tau}=h_2,\ldots,z^{(p)}_{q,\tau}=h_q,c^{(p)}_{\tau}=c\}$ and\par \quad\quad $n=\sum_{c=1}^{C}n_{i,h_2,\ldots,h_q}(c)$, where $i=1,\ldots,C$ and $h_j=1,\ldots,k_j$ for $j=2,\ldots,q$.
		\State \quad\quad Independently sample $\rho_s\sim \text{Geo}(1-\lambda_{i,h_2,\ldots,h_q}(i)), s=1,\ldots,n$.
		\State \quad\quad Sample $\boldsymbol{\lambda}_{i,h_2,\ldots,h_q}=\{\lambda_{i,h_2,\ldots,h_q}(1),\ldots,\lambda_{i,h_2,\ldots,h_q}(i),\ldots,\lambda_{i,h_2,\ldots,h_q}(C)\sim$\par
		\quad\quad $\text{Dir}\{\alpha\lambda_0(1)+n_{i,h_2,\ldots,h_q}(1),\ldots,\alpha\lambda_0(i)+\sum_{s=1}^{n}\rho_s,\ldots,\alpha\lambda_0(C)+n_{i,h_2,\ldots,h_q}(C)\}$.
		\State \quad Update $\boldsymbol{\bar\lambda_k}$: Compute $\boldsymbol{\bar\lambda_k}$ by Equation (\ref{self}).
		\State \quad Update $\boldsymbol{\lambda}_0$:
		\State \quad\quad For $r=1,\ldots,n_{i,h_2,\ldots,h_q}(c)$, sample $x_r\sim \text{Bernoulli}\left\{\frac{\alpha\lambda_0(c)}{r-1+\alpha\lambda_0(c)}\right\}$.
		\State \quad\quad Let $m_{i,h_2,\ldots,h_q}(c)=\sum_rx_r$ and $m_0(c)=\sum_{(i,h_2,\ldots,h_q)}m_{i,h_2,\ldots,h_q}(c)$.
		\State \quad\quad Sample $\boldsymbol{\lambda}_0=\{\lambda_0(1),\ldots,\lambda_0(C)\}\sim \text{Dir}\{\alpha_0/C+m_0(1),\ldots,\alpha_0/C+m_0(C)\}$.
		\State \quad Update
		$\{c^{(p)}_{\tau}:\tau=1,\ldots,S_p\}$, \(\{z^{(p)}_{j,\tau}: \tau=q+1,\ldots,S_p,j=1,\cdots,q\}\), $\boldsymbol{\theta}$, $\boldsymbol{k}$, $\tilde{\boldsymbol{\pi}}_{\boldsymbol{k}}$ as in \cite{sarkar2018bayesian}.
		\State \textbf{Stage 2} (Sample with determined ${\boldsymbol{k}}$):
		\State \quad Update $\boldsymbol{\pi}_{\boldsymbol{k}}$:
		\State \quad\quad Let $n_{j,w_j}(h_j)=\sum_p\sum_{\tau}1\{w^{(p)}_{j,\tau}=w_j, z^{(p)}_{j,\tau}=h_j\}$, where $w^{(p)}_{j,\tau}=c^{(p)}_{\tau-j}$.
		\State \quad\quad Sample $\boldsymbol{\pi}_{k_j}^{(j)}(w_j)=\{\pi_{1}^{(j)}(w_j),\ldots,\pi_{k_j}^{(j)}(w_j)\}\sim \text{Dir}\{\gamma_j+n_{j,w_j}(1),\ldots,\gamma_j+n_{j,w_j}(k_j)\}$.
		\State \quad Update $\boldsymbol{\lambda_k}$, $\boldsymbol{\bar\lambda}_{\boldsymbol{k}}$, $\boldsymbol{\lambda}_0$ as in \textbf{Stage 1}.
		\State \quad Update \(\boldsymbol{z}^{(p)}\): Sample from\par 
		\quad\quad\quad $p(z_{j,\tau}=h|z_{l,\tau}=h_l,l\neq j,\boldsymbol{\bar\lambda}_{\boldsymbol{k}},\boldsymbol{\pi_k},\boldsymbol{c})\propto\bar\lambda_{h_1,\ldots,h_{j-1},h,h_{j+1},\ldots,h_q}(c_{\tau})\pi^{(j)}_{h}(c_{\tau-j})$.
		\State \quad Update \(\boldsymbol{c}^{(p)}\): Sample from $p(c_{\tau}|\boldsymbol{\bar\lambda}_{\boldsymbol{k}},\boldsymbol{\pi_k},\boldsymbol{\theta},\boldsymbol{z})\propto\bar\lambda_{z_{1,\tau},z_{2,\tau},\ldots,z_{q,\tau}}(c_{\tau})f(\boldsymbol{y}_{t_{\tau}^1:t_{\tau}^2}|\boldsymbol{\theta}_{c_{\tau}})\prod_{j=1}^{q}\pi^{(j)}_{z_{j,\tau+j}}(c_{\tau})$.
		
		\State \quad Update $\boldsymbol{\theta}$ as in \textbf{Stage 1}.	
		\State \textbf{Update} $\alpha$: Compute likelihood value $L_{\alpha_{v}}$ from Equation (\ref{log-likeli}).
		\State \quad \textbf{if} $\min\left\{\frac{L_{\alpha_{v}}}{L_0},1\right\}>\text{rand}$ \textbf{then}
		$\alpha(v)\leftarrow\alpha_{v}$ and $L_0\leftarrow L_{\alpha_{v}}$.
		\State \quad \textbf{else}
		$\alpha(v)\leftarrow\alpha(v-1)$.
		\State \quad \textbf{end if}
		\EndFor
		\State \textbf{Determine $\alpha^*$}:
		\State \quad Use the average value of sampled $\alpha$ after burn-in as the most likely jump size threshold $\alpha^*$.
		\State \quad Given $\alpha^*$, repeat \textbf{Update segmentation}, \textbf{Stage 1} and \textbf{Stage 2} and collect final samples.
		\State \quad Compute the MLEs $\{\hat\xi_c\}$ for duration distribution using samples of \(\boldsymbol{c}^{(p)}\) and  \(\boldsymbol{d}^{(p)}\) for all $p$.
		\Ensure $\boldsymbol{k}$, $S^*_p$, $\{\hat\xi_c\}$, and samples of $\boldsymbol{\bar\lambda}_{\boldsymbol{k}}$, $\boldsymbol{\pi_k}$, $\boldsymbol{\theta}$.
	\end{algorithmic}
\end{algorithm}
\noindent conditioned on the learned model parameters $\boldsymbol{k}$, $\boldsymbol{\bar\lambda}_{\boldsymbol{k}}$, $\boldsymbol{\pi_k}$, $\boldsymbol{\theta}$ and duration times $\boldsymbol{d}$. We can directly use it for decoding the hidden super-states by sampling $\boldsymbol{c}$ and $\boldsymbol{z}$ from the posteriors. We need to first segment the observation sequence and then initialize the hidden super-state for each segment. Therefore, we use the same procedure described in Algorithm \ref{algorithm1} by sampling the jump size threshold and identifying change-points. Next, we initialize the allocation variables $\boldsymbol{z}$ based on the initialized $\boldsymbol{c}$ and the learned allocation distribution $\boldsymbol{\pi_k}$. Given the values of $\boldsymbol{k}$ and historical samples of $\boldsymbol{\bar\lambda}_{\boldsymbol{k}}$, $\boldsymbol{\theta}$ from the learned model, updating $\boldsymbol{c}$ and $\boldsymbol{z}$ is the same as in Algorithm \ref{algorithm1}. The collected samples of $\boldsymbol{c}$ are used to determine the hidden health states of this specific system by using the most persistent sample (i.e., the mode) for each segment. Figure \ref{decoding_state} provides an illustrative example. In Figure \ref{decoding_state}, we can see that state 1 appears most in the posterior samples for the first super-state $c_{1}$, and is therefore used as the estimated super-state for the first segment. The same selection criterion is used to determine the hidden super-states for all segments. Details of the decoding procedure are summarized in Algorithm \ref{algorithm 2}.
\begin{algorithm}[h]
	\caption{Decoding for HOHSMM}
	\label{algorithm 2}
	\begin{algorithmic}[1]
		\Require  \(\boldsymbol{y}\), sample size $l$, and learned model parameters $\alpha_{\text{min}}$, $\alpha_{\text{max}}$, $\boldsymbol{k}$, $\boldsymbol{\bar\lambda}_{\boldsymbol{k}}$, $\boldsymbol{\pi_k}$, $\boldsymbol{\theta}$.
		\State \textbf{Initialization}: Set the initial likelihood value: $L_0\leftarrow e^{-10^{10}}$.
		\For{$v=1$ to $l$}
		
		\State Sample $\alpha_{v}\sim U(\alpha_{\text{min}},\alpha_{\text{max}})$.
		\State \textbf{Update segmentation}:
		\State \quad Compute distances between two adjacent data points in $\boldsymbol{y}$ and identify change-points\par 
		\quad given threshold $\alpha_{v}$. Compute the center of the observed data points for each segment \par 
		\quad and the initialize hidden super-state $c_{\tau}$ by determining the clustering label based on the\par 
		\quad learned clustering rules. Merge adjacent segments that have the same label and derive\par
		\quad the number of segments $S$ and duration times $\boldsymbol{d}$, where $\tau=1,\ldots,S$.
		\State \textbf{Decode}: 
		\State \quad \textbf{Initialize} $\boldsymbol{z}$:
		\State \quad\quad Sample $(z_{j,\tau}|c_{\tau-j})\sim \text{Mult}(\{1,\ldots,k_j\},\pi^{(j)}_1(c_{\tau-j}),\ldots,\pi^{(j)}_{k_j}(c_{\tau-j}))$, where $j=1,\ldots,q$.
		\State \quad\textbf{Update} $\boldsymbol{c}$: 
		\State \quad\quad Sample $c_{\tau}$ from $p(c_{\tau}|\boldsymbol{\bar\lambda}_{\boldsymbol{k}},\boldsymbol{\pi_k},\boldsymbol{\theta},\boldsymbol{z})\propto\bar\lambda_{z_{1,\tau},z_{2,\tau},\ldots,z_{q,\tau}}(c_{\tau})f(\boldsymbol{y}_{t_{\tau}^1:t_{\tau}^2}|\boldsymbol{\theta}_{c_{\tau}})\prod_{j=1}^{q}\pi^{(j)}_{z_{j,\tau+j}}(c_{\tau})$.
		\State \quad\textbf{Update} $\boldsymbol{z}$:
		\State \quad\quad Sample $z_{j,\tau}$ from \par \quad\quad\quad\quad$p(z_{j,\tau}=h|z_{l,\tau}=h_l,l\neq j,\boldsymbol{\bar\lambda}_{\boldsymbol{k}},\boldsymbol{\pi_k},\boldsymbol{c})\propto\bar\lambda_{h_1,\ldots,h_{j-1},h,h_{j+1},\ldots,h_q}(c_{\tau})\pi^{(j)}_{h}(c_{\tau-j})$.
		\State \textbf{Update $\alpha$}: Compute the likelihood value $L_{\alpha_{v}}$ from Equation (\ref{log-likeli}). 
		\State \quad \textbf{if}  $\min\left\{\frac{L_{\alpha_{v}}}{L_0},1\right\}>\text{rand}$ \textbf{then} $\alpha(v)\leftarrow\alpha_{v}$ and $L_0\leftarrow L_{\alpha_{v}}$.
		\State \quad \textbf{else} $\alpha(v)\leftarrow\alpha(v-1)$.
		\State \quad \textbf{end if}
		\EndFor
		\State \textbf{Determine $\alpha^*$}:
		\State \quad Use the average value of sampled $\alpha$ after burn-in as the most likely jump size threshold $\alpha^*$.
		\State \quad Given $\alpha^*$, repeat \textbf{Update segmentation} and \textbf{Decode} for final samples of $c^*_{\tau}$, $\tau=1,\ldots,S^*$.
		\State \textbf{Determine} hidden health states $\{\tilde{c}_{\tau}:\tau=1,\ldots,S^*\}$ by using the most persistent samples.
		\Ensure $\{\tilde{c}_{\tau}:\tau=1,\ldots,S^*\}$.
	\end{algorithmic}
\end{algorithm}
\section{RUL Estimation}
In this section, we estimate the RUL given the decoded hidden health states. For notational convenience, we omit the superscript of the number of segments $S^*$ in the following analysis. It is impossible to analytically compute the RUL in an HOHSMM due to higher order state transitions. Therefore, we use a simulation approach to predict the RUL, which is the expected time from the current health state to the failure state. Before presenting the RUL estimation method, we first show how to identify the failure state. The HOHSMM is trained using multiple run-to-failure independent observation sequences and historical samples of hidden super-states can be used to identify the failure state. For each sequence, from the hidden super-state samples $\{c^{(p)}_{\tau}\}$, we identify the failure state $c^{(p)}_{F}$ by choosing the most persistent state in the last $f$ states \cite{tobon2012data},
\begin{equation}
\begin{aligned}
&\text{Super-state sequence }=\left(c^{(p)}_1,c^{(p)}_2,\cdots,c^{(p)}_{S_p}\right),\\
&\text{Last $f$ states }=\left(c^{(p)}_{S_p-f+1},\cdots,c^{(p)}_{S_p-2},c^{(p)}_{S_p-1},c^{(p)}_{S_p}\right),
\end{aligned}
\label{failure}
\end{equation}
The value of $f$ can be chosen based on experience. Then, the final failure state $c_{F}$ is given as the most persistent state in all $c^{(p)}_{F}$, $p=1,\ldots,P$. 
\begin{figure}[h]
	\centering
	\includegraphics[width=4.3in]{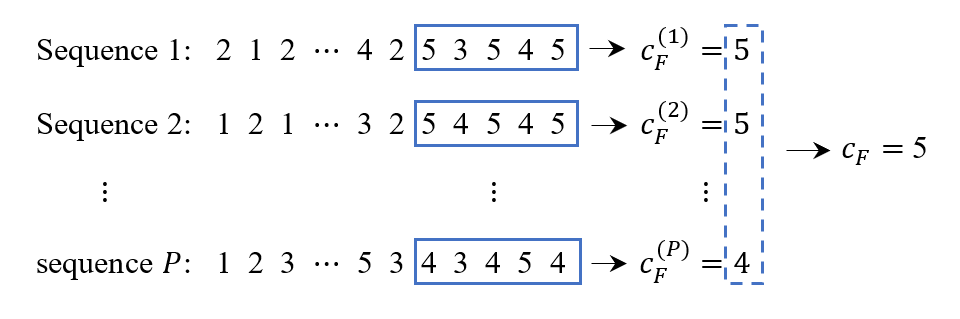}
	\caption{Illustration for identifying the failure state $c_{F}$ with $f=5$}
	\label{failure_state}
\end{figure}
Figure \ref{failure_state} illustrates the procedure to select the final failure state. For illustrative purpose, we arbitrarily use the last five super-states to identify the failure state for each sequence by choosing the most persistent state appeared in the last five states. We can see that state 5 appears most in the last five states for sequence 1 and 2, and state 4 is the most persistent state in the last five states for sequence $P$. We then choose the most persistent state in the identified failure states of all sequences as the final failure state, which is state 5 in this example.

Given the decoded hidden health states $\{\tilde{c}_{\tau}:\tau=1,\ldots,S\}$ and the identified failure state $c_{F}$, we use a simulation method to estimate the RUL. We simulate $M$ hidden super-state paths. Each path starts from states $(\tilde{c}_{S-q+1},\ldots,\tilde{c}_{S})$ and the next state $c_{S+1}$ is generated by drawing a sample from the multinomial distribution with probabilities $\left(p(c_{S+1}=1|\tilde{\boldsymbol{c}}_{(S-q+1):S}),\ldots,p(c_{S+1}=C|\tilde{\boldsymbol{c}}_{(S-q+1):S})\right)$, which are computed using Equation (\ref{transitions}) given the learned parameters $\boldsymbol{\bar\lambda}_{\boldsymbol{k}}$ and $\boldsymbol{\pi_k}$. Repeat this procedure by considering $c_{S+1}$ as the current health state until the failure state $c_{F}$ is first reached. Denote the $i^{th}$ paths by $\{c_{S+1},\ldots,c_{S+N_i}\}$, where $N_i$ is the total number of super-states generated in the $i^{th}$ paths and we have $c_{S+N_i}=c_{F}$ for all $i=1,\ldots,M$. We estimate the mean RUL by the following equations
\begin{equation}
RUL^i=\sum_{j=1}^{N_i}D_{c_{S+j}},
\end{equation}
\begin{equation}
\label{RUL}
RUL=\frac{1}{M}\sum_{i=1}^{M}RUL^i,
\end{equation}
where the mean duration time $D_c$ for state $c$ is computed from the estimated duration distribution $g(d;\hat\xi_{c})$ (e.g., $D_c=\hat\xi_{c}$ for Poisson distributions) and $c=1,\ldots,C$. Details of estimating RUL are summarized in Algorithm \ref{algorithm 3}.

\begin{algorithm}[H]
	\caption{RUL estimation}
	\label{algorithm 3}
	\begin{algorithmic}[1]
		\Require  Learned model parameters $\boldsymbol{\bar\lambda}_{\boldsymbol{k}}$, $\boldsymbol{\pi_k}$, $\{\hat\xi_c\}$, decoded hidden super-state sequence $\{\tilde{c}_{\tau}:\tau=1,\ldots,S\}$, failure state $c_{F}$, and the number of simulation paths $M$.
		\State \textbf{Compute transition probability matrix}: For each possible combination of $\boldsymbol{c}_{(\tau-q):(\tau-1)} $,\par  
		compute $p(c_{\tau}=c|\boldsymbol{c}_{(\tau-q):(\tau-1)})$ using Equation (\ref{transitions}), $c=1,\ldots,C$.
		\State \textbf{Compute mean duration time}: $D_c$, $c=1,\ldots,C$.
		\For{$i=1$ to $M$}
		\State \textbf{Initialize}: $\boldsymbol{c}_{\text{now}}\leftarrow(\tilde{c}_{S-q+1},\ldots,\tilde{c}_{S})$, $N_i\leftarrow0$, $RUL^i\leftarrow0$.
		\While{$\boldsymbol{c}_{\text{now}}(q)\neq c_F$}
		\State Sample $c\sim \text{Mult}(\{1,\ldots,C\},p(c=1|\boldsymbol{c}_{\text{now}}),\ldots,p(c=C|\boldsymbol{c}_{\text{now}}))$.
		\State $N_i\leftarrow N_i+1$.
		\State $RUL^i\leftarrow RUL^i+D_c$.
		\State $\boldsymbol{c}_{\text{now}}(1:(q-1))\leftarrow\boldsymbol{c}_{\text{now}}(2:q)$ and $\boldsymbol{c}_{\text{now}}(q)\leftarrow c$.
		\EndWhile
		\EndFor
		\State \textbf{Compute mean RUL}: $RUL=\frac{1}{M}\sum_{i=1}^{M}RUL^i$.
		
		\Ensure $RUL$.
	\end{algorithmic}
\end{algorithm}
\section{Simulation Experiment}
We design the following simulation experiment to evaluate the performance of our proposed sampling method for the HOHSMM with consideration of multiple independent observation sequences.
\subsection{Model setting}
Consider a 3-order case with state space \(\mathscr{C}=\{1,2,3\}\), and set $\{c_{\tau-1},c_{\tau-2},c_{\tau-3}\}$ as the important lags. We independently generate three observation sequences. The sample size $T_p$ for each sequence is randomly chosen from 800 to 1000, $p=1,2,3$. The true transition probability tensors $\boldsymbol{\lambda}_{h_1,h_2,h_3}$ are generated as follows,
\begin{equation*}
\lambda_{h_1,h_2,h_3}(1)=\frac{u_1^2}{u_1^2+(1-u_1)^2},\quad u_1\sim U(0,1),
\end{equation*}
\begin{equation*}
\lambda_{h_1,h_2,h_3}(2)=\frac{u_2^2}{u_2^2+(1-u_2)^2}[1-\lambda_{h_1,h_2,h_3}(1)],\quad u_2\sim U(0,1),
\end{equation*}
\begin{equation*}
\lambda_{h_1,h_2,h_3}(3)=1-\lambda_{h_1,h_2,h_3}(1)-\lambda_{h_1,h_2,h_3}(2),
\end{equation*}
where $h_1,h_2,h_3\in\{1,2,3\}$. By excluding self-transitions, we obtain

\begin{equation*}
\bar\lambda_{i,h_2,h_3}(j)=\frac{\lambda_{i,h_2,h_3}(j)(1-\delta_{ij})}{1-\lambda_{i,h_2,h_3}(i)},\quad \delta_{ij}=\begin{cases}
1 &\text{if $i=j$,}\\
0 &\text{otherwise.}
\end{cases}
\end{equation*}
The hyper-parameters in the priors are set as $\alpha_0=1$ and $\gamma_j=1/C=1/3$ for all $j$. We consider normal emission distribution  $f(y|c_{\tau}=c)=\text{Normal}(y|\mu_c,\sigma_c^2)$ with parameters $\mu_c=-3,0,3$ for $c=1,2,3$, respectively and $\sigma^2_c=0.5^2$ for all $c$ \cite{sarkar2018bayesian}. The duration time is assumed to follow a Poisson distribution $g(d|\xi_c)$. The intensities $\xi_c$ are $15,10,5$ for $c=1,2,3$, respectively.

\subsection{Results}

Based on the three simulated observation sequences, the support for jump size threshold $\alpha$ is $(0.052,3.204)$. After $l=50$ iterations, we obtain the most likely jump size threshold $\alpha^*=1.726$. The three observation sequences are segmented based on $\alpha^*$ with the total numbers of segments $S_1=77$, $S_2=78$, $S_3=77$, which are close to the true numbers $S^{\text{true}}_1=82$, $S^{\text{true}}_2=78$, $S^{\text{true}}_3=82$.

\begin{figure}[h]
	\centering
	\includegraphics[width=2.9in]{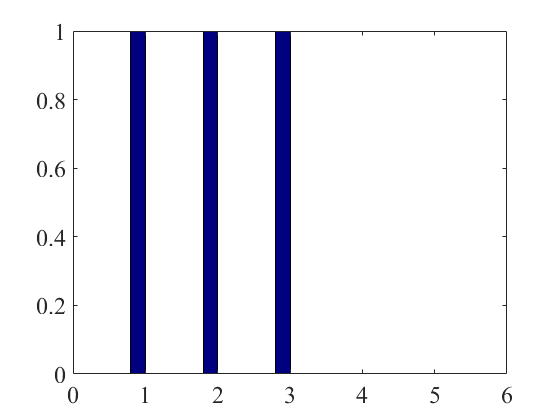}
	\caption{HOHSMM simulation result: The inclusion probability given $\alpha^*$}
	\label{inlusion}
\end{figure}

Important lags inclusion result given $\alpha^*$ is shown in Figure \ref{inlusion}. Given the collected samples of ${\boldsymbol{k}}$, the inclusion probability for each lag is derived as the proportion of $k_j>1$. From Figure \ref{inlusion}, we can see that the important lags are identified correctly since the true important lags are set as $\{c_{\tau-1},c_{\tau-2},c_{\tau-3}\}$. Table \ref{HOHSMM_estimated_paramters} presents the estimated parameters. Given the collected samples of $\mu_c$ and $\sigma_c$ for the emission distributions, each parameter is estimated by computing the average value of the respective samples, where $c=1,2,3$. The MLEs of intensities for duration time distributions are directly derived based on the final duration time samples. In this case, the MLE $\hat\xi_c$ is computed by using the sample mean of all duration times staying in state $c$. From Table \ref{HOHSMM_estimated_paramters}, we can see the estimated parameters are close to the true values.

\begin{figure}[H]
	\centering
	\includegraphics[width=6in]{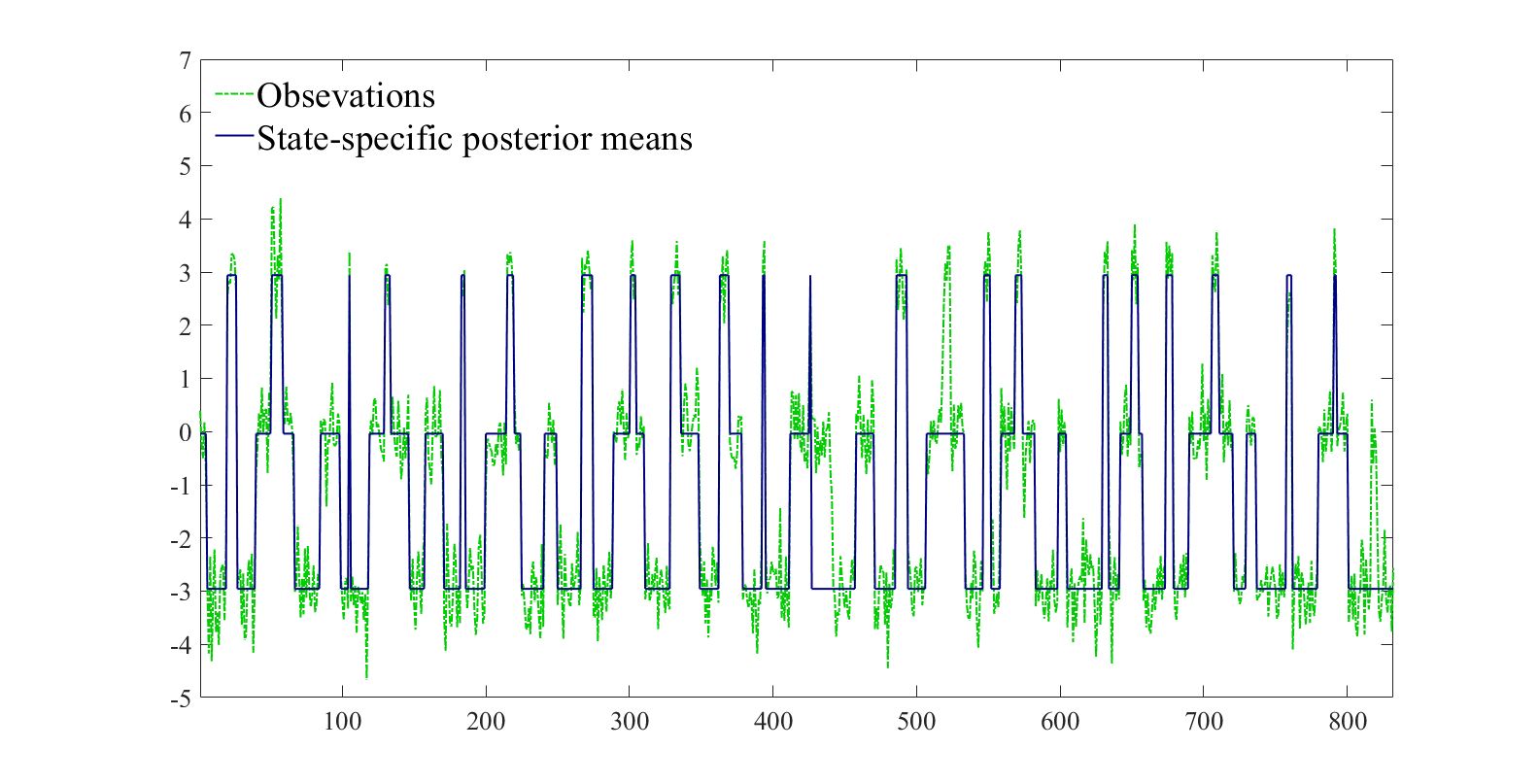}
	\caption{HOHSMM simulation result: The state-specific posterior means (blue solid line) super-imposed over one observation sequence (green dashed line)}
	\label{posterior_mean}
\end{figure}

\begin{table}[h]
	\centering
	\begin{tabular}{c|c|c|c}
		\hline
		Super hidden-state $c$ & $\hat\mu_c$   & $\hat\sigma_c$  & $\hat\xi_c$ \\ \hline
		1 & -2.96 & 0.36 & 14.99 \\ \hline
		2 & -0.04  & 0.47 & 10.77  \\ \hline
		3 & 2.95 & 0.36 & 4.70  \\ \hline
	\end{tabular}
	\caption{HOHSMM simulation result: Estimated parameters given $\alpha^*$}
	\label{HOHSMM_estimated_paramters}
\end{table}

We randomly select one sequence to evaluate the accuracy of decoding. We compare the state-specific posterior means of the emission distributions and the observation sequence (illustrated in Figure \ref{posterior_mean}). From Figure \ref{posterior_mean}, we can see the estimated emission distributions based on the decoded hidden super-states describe the observations well. The proposed sampling method is effective for HOHSMM inference in the simulation experiment.

\section{Case Study: Turbofan Engines Prognostics Analysis}
To further demonstrate the practical utility of the proposed HOHSMM on diagnostics and prognostics, we conduct a case study on turbofan engines from the NASA Prognostic Data Repository. The C-MAPSS dataset is used in this paper, which is generated using a model-based simulation program (named Commercial Modular Aero-Propulsion System Simulation) developed by NASA \cite{saxena2008damage}. 

For illustrative purpose, only the training set in dataset FD001 is used in this paper, which contains 100 engines' run-to-failure trajectories. All trajectories in this training set are simulated under the same operational condition and have only one fault mode caused by High Pressure Compressor (HPC) degradation \cite{frederick2007user}. Each trajectory is recorded in a given operational cycle, consisting of three values for operational settings and 21 values for engine performance sensor measurements. We randomly choose six trajectories to train the HOHSMM and randomly choose another four trajectories for testing.

Multiple sensor measurements bring dimensionality challenge for data analysis. To keep effective discriminant information and eliminate the redundant one, feature fusion process is used to transfer a set of sensors to a single health indicator. To obtain the health indicator, we use principle component analysis (PCA), which is an efficient technique in compressing information and eliminating the correlations between variables. The first principle component (FPC), accounting for the largest variability in data, is used as the health condition indicator \cite{moghaddass2014integrated}. In the HOHSMM, we assume that the health indicator (i.e., FPC) follows a state-specific normal distribution. We assume there are seven health states since it has been shown that the hidden health conditions are well represented by seven states \cite{moghaddass2014integrated}. 

\begin{figure}[h]
	\centering
	\includegraphics[width=2.9in]{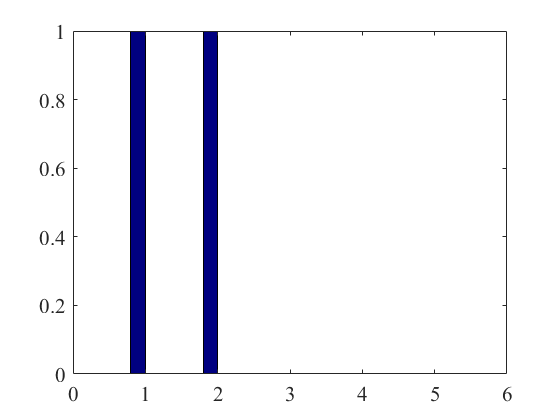}
	\caption{Case study result: The inclusion probability given $\alpha^*$}
	\label{inlusion_nasa}
\end{figure}

\begin{figure}[h]
	\centering
	\includegraphics[width=5.5in]{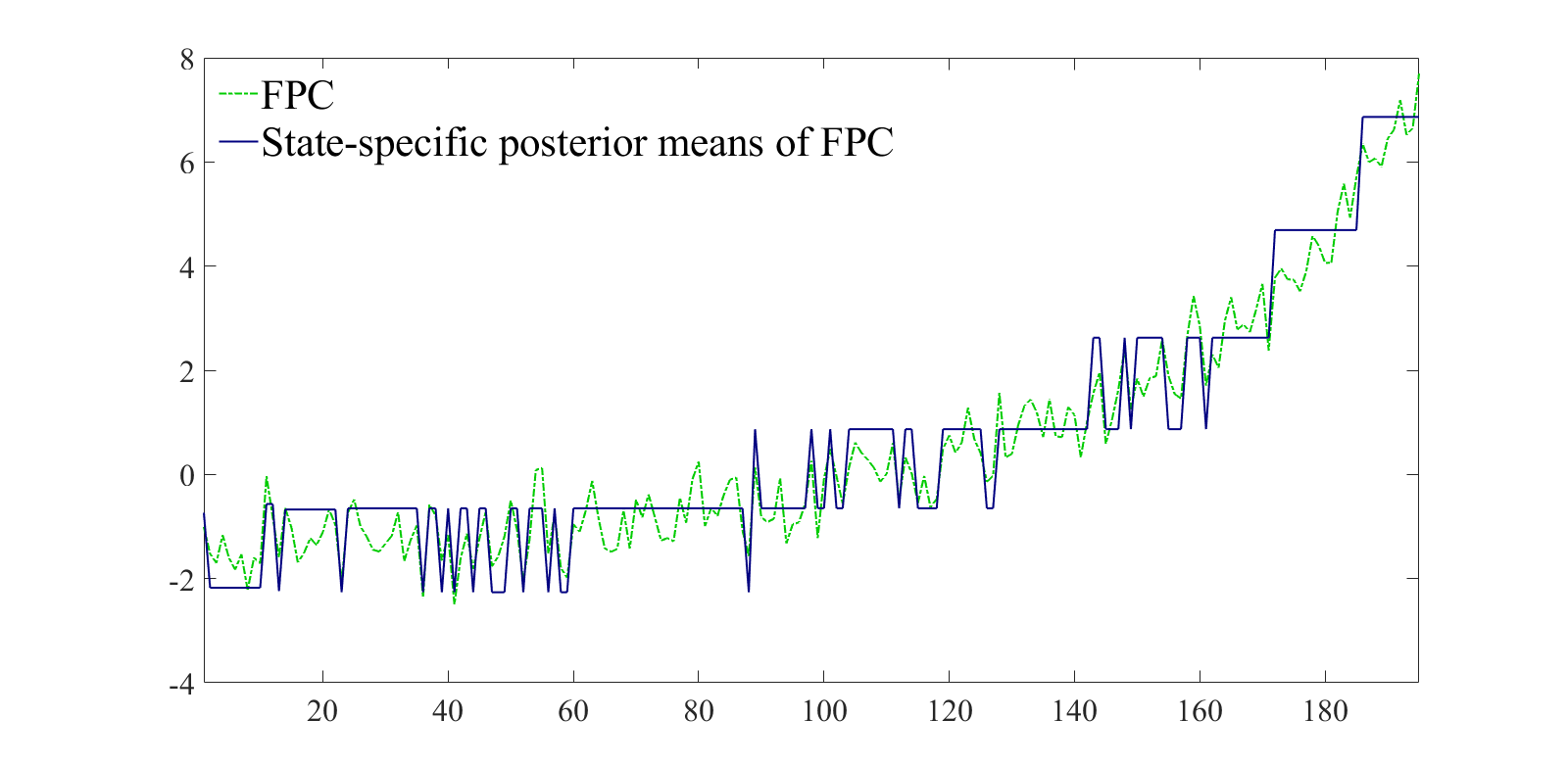}
	\caption{Case study result: The state-specific posterior means of FPC (blue solid line) super-imposed over the FPC sequence (green dashed line) for training unit 1}
	\label{posterior_mean_nasa}
\end{figure}

From the important lags inclusion result (shown in Figure \ref{inlusion_nasa}), 
we can see that the hidden health state sequence is governed by a second order Markov chain, implying that the health state transition of turbofan engines depends on its past two history states. The performance of hidden state decoding on training data is illustrated in Figure \ref{posterior_mean_nasa} using the first training trajectory as an example. We compare the state-specific posterior means of FPC and the true FPC computed from raw sensor data. We can see the decoding performance is very good since the estimated emission distributions based on the decoded hidden super-state sequence describe the computed FPC well.

\begin{figure}[H]
	\centering
	\subfigure[Training unit 1]{
		\begin{minipage}{2.9in}
			\centering
			\includegraphics[width=2.9in]{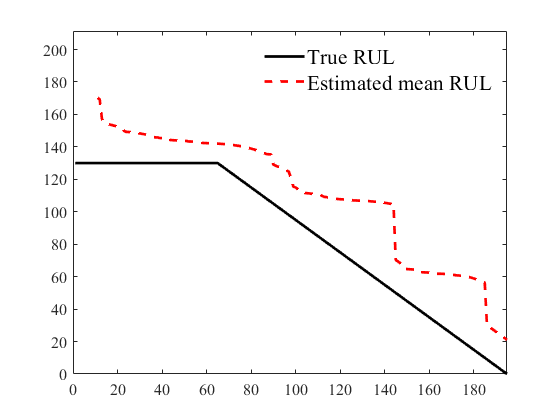}
		\end{minipage}
	}
	\subfigure[Training unit 2]{
		\begin{minipage}{2.9in}
			\centering
			\includegraphics[width=2.9in]{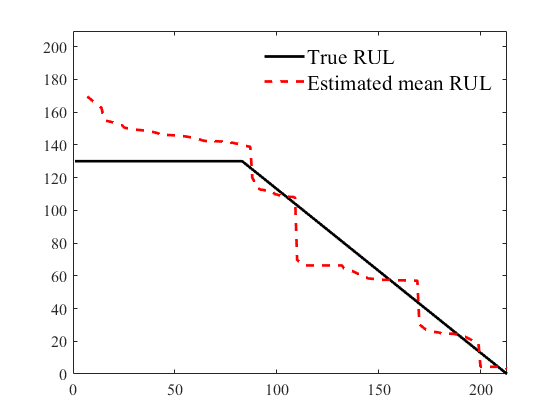}
		\end{minipage}
	}
	\subfigure[Training unit 3]{
		\begin{minipage}{2.9in}
			\centering
			\includegraphics[width=2.9in]{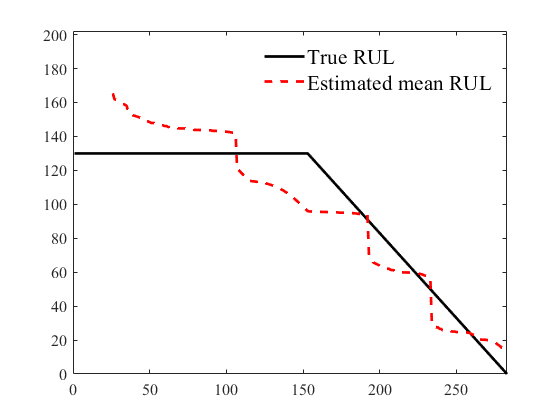}
		\end{minipage}
	}
	\subfigure[Training unit 4]{
		\begin{minipage}{2.9in}
			\centering
			\includegraphics[width=2.9in]{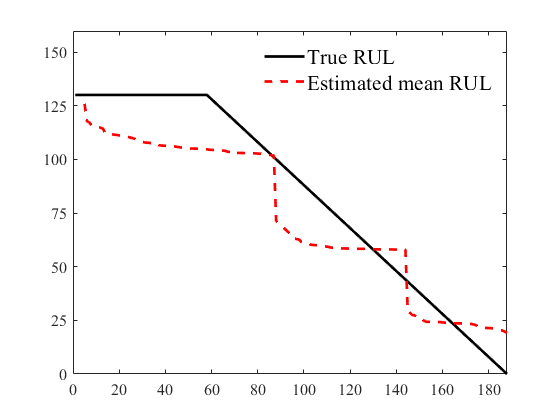}
		\end{minipage}
	}
	\subfigure[Training unit 5]{
		\begin{minipage}{2.9in}
			\centering
			\includegraphics[width=2.9in]{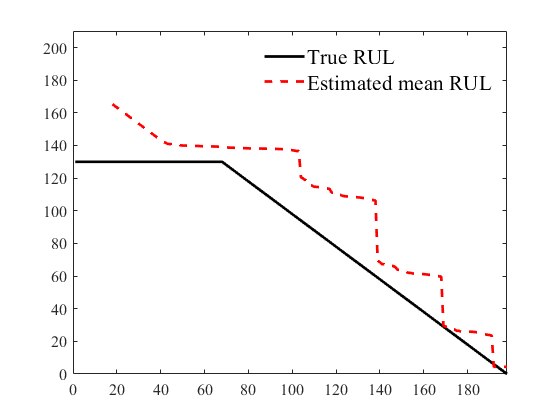}
		\end{minipage}
	}
	\subfigure[Training unit 6]{
		\begin{minipage}{2.9in}
			\centering
			\includegraphics[width=2.9in]{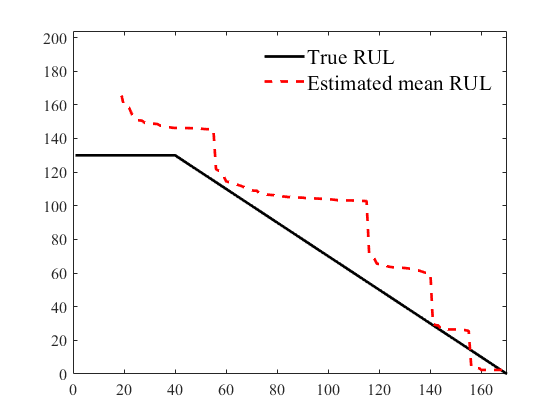}
		\end{minipage}
	}
	\caption{Case study result: RUL prediction for training units $1-6$}
	\label{RUL_train}
	
\end{figure}

Next, we use the learned HOHSMM to predict the RULs for both training and testing units using the simulation method presented in Section 5. For each unit, we generate 100 hidden super-state paths and compute the mean RUL. Since the degradation in a system is generally not noticeable after the unit has been operated for some period of time, it is reasonable to estimate the RUL using a piece-wise linear function \cite{heimes2008recurrent}, which limits the maximum value of the RUL. Thus, a piece-wise RUL plot is used to represent the true RUL, which serves as the benchmark for the predicted RUL. In Figure \ref{RUL_train}, we present RUL estimation results of the six training units. We can see the estimated mean RUL is close to the true RUL. 

\begin{figure}[H]
	\centering
	\subfigure[Testing unit 1]{
		\begin{minipage}{2.1in}
			\centering
			\includegraphics[width=2.1in]{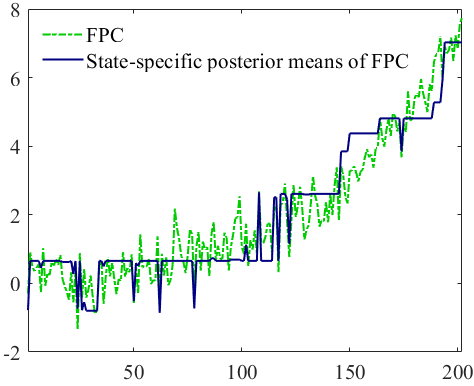}
		\end{minipage}
	}
	\subfigure[Testing unit 2]{
		\begin{minipage}{2.1in}
			\centering
			\includegraphics[width=2.1in]{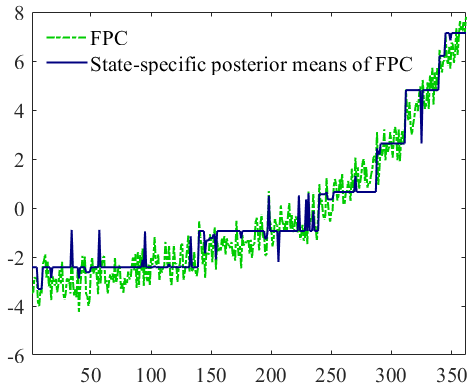}
		\end{minipage}
	}
	\subfigure[Testing unit 3]{
		\begin{minipage}{2.1in}
			\centering
			\includegraphics[width=2.1in]{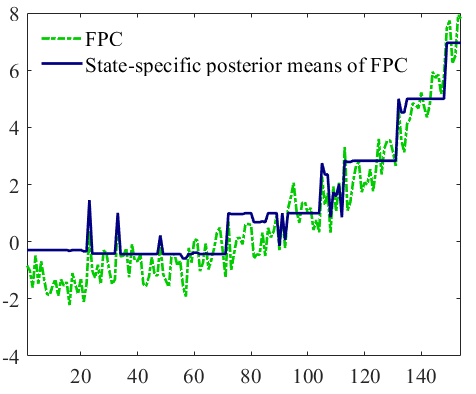}
		\end{minipage}
	}
	\subfigure[Testing unit 4]{
		\begin{minipage}{2.1in}
			\centering
			\includegraphics[width=2.1in]{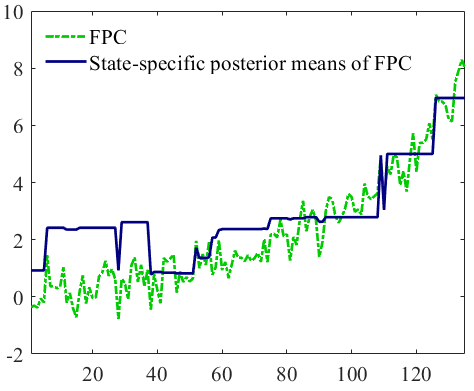}
		\end{minipage}
	}
	
	\caption{Case study result: Hidden health state decoding for testing units $1-4$}
	\label{decode_test}
	
\end{figure}

\begin{figure}[H]
	\centering
	\subfigure[Testing unit 1]{
		\begin{minipage}{2.1in}
			\centering
			\includegraphics[width=2.1in]{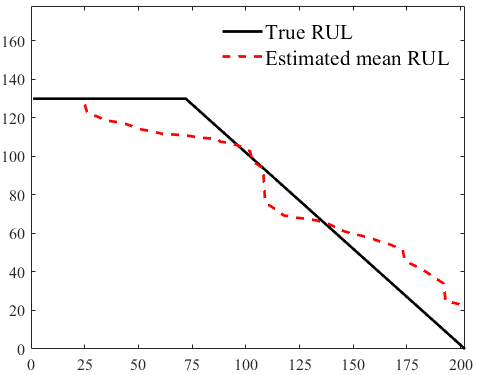}
		\end{minipage}
	}
	\subfigure[Testing unit 2]{
		\begin{minipage}{2.1in}
			\centering
			\includegraphics[width=2.1in]{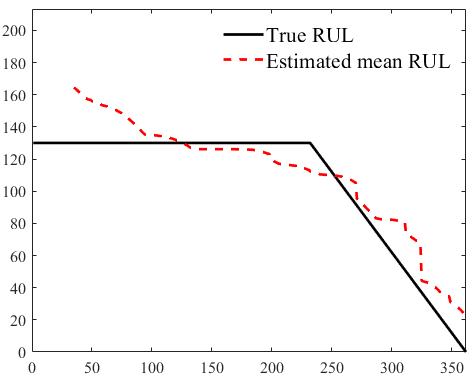}
		\end{minipage}
	}
	\subfigure[Testing unit 3]{
		\begin{minipage}{2.1in}
			\centering
			\includegraphics[width=2.1in]{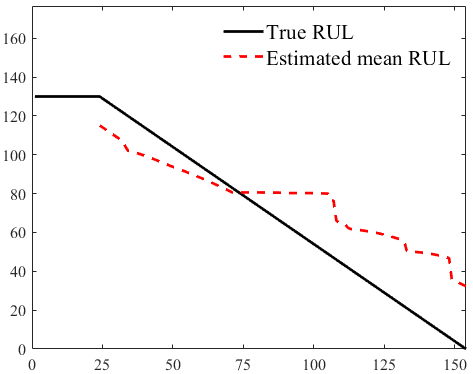}
		\end{minipage}
	}
\subfigure[Testing unit 4]{
	\begin{minipage}{2.1in}
		\centering
		\includegraphics[width=2.1in]{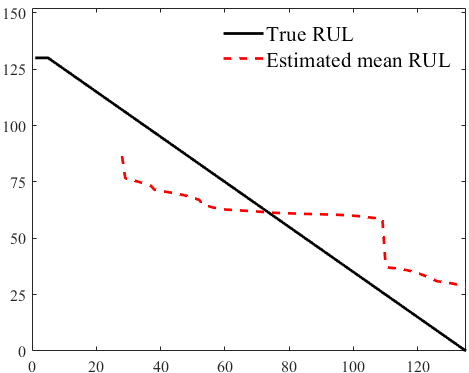}
	\end{minipage}
}
	
	\caption{Case study result: RUL prediction for testing units $1-4$}
	\label{RUL_test}
	
\end{figure}

We further use four different units to test the diagnostics and prognostics performance of the learned model. First, we compute the FPC for the four testing units based on the PCA results obtained from the training data. The hidden health states are decoded using Algorithm \ref{algorithm 2} and the decoding results are presented in Figure \ref{decode_test}. We can see that the decoding performance of the testing units is good since the state-specific means of FPC can well describe the computed FPC. By generating 100 paths for each test unit, we obtain the estimated mean RUL (shown in Figure \ref{RUL_test}). The estimation results are acceptable for the testing units since the estimated mean RUL is close to the true RUL. 


\section{Conclusion}
In this paper, we consider the problem of decoding the hidden health states and predicting the RUL for systems with unobservable health conditions and complex transition dynamics based on observations. We develop a flexible prognostics framework based on an HOHSMM. Our framework is flexible in that the HOHSMM allows the hidden state to depend on its more distant history instead of only depending on the current state and assumes generally distributed state duration. The proposed HOHSMM includes the HMM and HSMM as two special cases. A Gibbs sampling algorithm is designed for HOHSMM inference and is evaluated by conducting a simulation experiment. The results show that the proposed HOHSMM sampler is effective for learning model parameters from the observed data. Given the learned model, a decoding algorithm is developed to assess the current hidden health state of a functioning system in operation. The RUL is then predicted using a simulation approach by generating hidden state sequences from the current state to the failure state. The NASA turbofan engine dataset (i.e., C-MAPSS dataset) is used to demonstrate the practical utility of the proposed prognostics framework. Our case study shows that the HOHSMM-based prognostics framework provides satisfactory hidden health state assessment and RUL estimation for complex systems.

The framework presented in this paper has raised a few important questions that require further study. First, the state space is generally unknown and the true number of states also need to be learned from the observed data. The existing hierarchical Dirichlet process HMM (HDP-HMM) provides a powerful framework for inferring arbitrarily large state complexity from data \cite{Hierarchical2006}. Moreover, the hierarchical Dirichlet process HSMM (HDP-HSMM) allows for both Bayesian nonparametric inference of state complexity as well as general duration distributions \cite{johnson2013bayesian}. A promising direction for future research is to consider a more general model, the hierarchical Dirichlet process HOHSMM, to address the unknown state space issue in our proposed HOHSMM-based prognostics framework. Second, there generally exists heterogeneity among different operating systems (or components), even in the same environmental conditions. It is also necessary to extend our prognostics framework to account for the unit-to-unit differences in the future work.

\clearpage
\bibliographystyle{ieeetr}
\bibliography{bibfile}
\end{document}